\documentclass[pdflatex,sn-basic]{sn-jnl}

\usepackage{graphicx}%
\usepackage{multirow}%
\usepackage{amsmath,amssymb,amsfonts}%
\usepackage{amsthm}%
\usepackage{mathrsfs}%
\usepackage[title]{appendix}%
\usepackage{xcolor}%
\usepackage{textcomp}%
\usepackage{manyfoot}%
\usepackage{booktabs}%
\usepackage{algorithm}%
\usepackage{algorithmicx}%
\usepackage{algpseudocode}%
\usepackage{listings}%
\usepackage{siunitx}
\usepackage{subfigure}
\usepackage[dvipsnames]{xcolor}
\usepackage{placeins}
\usepackage{adjustbox}

\usepackage[version=4]{mhchem}

\DeclareFontFamily{U}{mathb}{\hyphenchar\font45}
\DeclareFontShape{U}{mathb}{m}{n}{
      <5> <6> <7> <8> <9> <10> gen * mathb
      <10.95> mathb10 <12> <14.4> <17.28> <20.74> <24.88> mathb12
      }{}
\DeclareSymbolFont{mathb}{U}{mathb}{m}{n}
\DeclareMathSymbol{\Earth}{3}{mathb}{"43}
\newcommand{\RE}{R$_{\rm \Earth}$}
\newcommand{\ME}{M$_{\rm \Earth}$}

\begin{document}

\title[Interior structure]{Fundamentals of interior modelling and challenges in the interpretation of observed rocky exoplanets}

\author*[1]{\fnm{Philipp} \sur{Baumeister}}\email{philipp.baumeister@fu-berlin.de}
\author*[2]{\fnm{Francesca} \sur{Miozzi}}\email{fmiozzi@carnegiescience.edu}
\author[3]{\fnm{Claire Marie} \sur{Guimond}}
\author[4]{\fnm{Marie-Luise} \sur{Steinmeyer}}
\author[4]{\fnm{Caroline} \sur{Dorn}}
\author[5]{\fnm{Shun-Ichiro} \sur{Karato}}
\author[6]{\fnm{Emeline} \sur{Bolmont}}
\author[6]{\fnm{Alexandre} \sur{Revol}}
\author[1]{\fnm{Alexander} \sur{Thamm}}
\author[1]{\fnm{Lena} \sur{Noack}}

\affil[1]{\orgdiv{Department of Earth Sciences}, \orgname{Freie Universität Berlin}, \orgaddress{\street{Malteserstrasse 74-100}, \postcode{12249} \city{Berlin}, \country{Germany}}}
\affil[2]{\orgdiv{Earth and Planets Laboratory}, \orgname{Carnegie Institution for Science}, \orgaddress{\street{5241 Broad Branch Road NW}, \postcode{20015}, \city{Washington D.C.},\country{United States}}}
\affil[3]{\orgdiv{Atmospheric, Oceanic, and Planetary Physics, Department of Physics}, \orgname{University of Oxford}, \orgaddress{\street{Parks Rd}, \city{Oxford} \postcode{OX1 3PU}, \country{United Kingdom}}}
\affil[4]{\orgdiv{Institute for Particle Physics and Astrophysics}, \orgname{ETH Zurich}, \orgaddress{\street{Wolfgang-Pauli-Strasse 27}, \postcode{8093} \city{Zürich}, \country{Switzerland}}}
\affil[5]{\orgdiv{Department of Earth and Planetary Sciences}, \orgname{Yale University}, \orgaddress{\street{210 Whitney Ave}}, \city{New Haven}, \postcode{CT 06520}, \country{United States}}
\affil[6]{\orgdiv{Department of Astronomy}, \orgname{Geneva University}, \orgaddress{\street{51 Chemin Pegasi}, \city{Sauverny}, \postcode{1290}, \country{Switzerland}}}


\abstract{Most our knowledge about rocky exoplanets is based on their measure of mass and radius. These two parameters are routinely measured and are used to categorise different populations of observed exoplanets. They are also tightly linked to the planet's properties, in particular those of the interior. As such they offer the unique opportunity to interpret the observations and potentially infer the planet's chemistry and structure. Required for the interpretation are models of planetary interiors, calculated a priori, constrained using other available data, and based on the physiochemical properties of mineralogical phases. This article offers an overview of the current knowledge about exoplanet interiors, the fundamental aspects and tools for interior modelling and how to improve the contraints on the models, along with a discussion on the sources of uncertainty.
The origin and fate of volatiles, and their role in planetary evolution is discussed.\\The chemistry and structure of planetary interiors have a pivotal role in the thermal evolution of planets and the development of large scale properties that might become observables with future space missions and ground-based surveys. As such, having reliable and well constrained interior models is of the utmost importance for the advancement of the field. 
}


\maketitle

\section{Introduction}
Nearly 6000 exoplanets have been detected so far\footnote{\url{https://exoplanetarchive.ipac.caltech.edu/}}, and this number continues to grow steadily. 
Although a variety of techniques exist for exoplanet detection, almost all low-mass planets have been discovered via the transit method, using data from the Kepler \citep{Borucki2010} and TESS missions \citep{Ricker2015}, and via the radial velocity (RV) method \citep[e.g.][]{Vogt_HIRES_1994,Mayor2011,Seifahrt_MaroonX_2018,Pepe2021}. 
The upcoming PLATO mission \citep{Rauer2014, rauer2025PLATOMission}, scheduled for launch in 2026, is expected to significantly increase the number of detected exoplanets. In addition, the Nancy Roman Space Telescope will be the first space mission with designated capabilities for direct imaging of exoplanets \citep{spergel2015WideFieldInfrarRed, akeson2019WideField, carrion-gonzalez2021CatalogueExoplanets}.
Kepler, TESS and PLATO rely on the transit method for planet detection, which is inherently biased towards small stars, large planets, and planets in close orbits. The large majority of small planets ($<$ 1.5 \RE) have periods below 100 days \citep[e.g.][]{2018AJ....156..264F}. Detecting rocky worlds, especially around Sun-like stars, remains particularly challenging. The number of detected potentially-rocky planets with measured mass and radius — on which this article focuses — is currently limited to around $\sim$100.

\subsection{Observed parameters and uncertainties}
The key characteristics of exoplanets used to infer their interior structure and composition are their mass, measured with RV or transit timing variations (TTV), and radius, measured with transit photometry. However, the uncertainties in these measurements are strongly dependent on the precision of stellar properties. Stellar parameters impose strict limits on the achievable precision, with the best uncertainties being approximately 2\% for stellar radius and 5--10\% for stellar mass \citep[e.g.,][]{tayar2022GuideRealistic}. These values depend on the stellar type, age, and the detection method used. For young stars, typical uncertainties are larger, in the range of 10--20\%. PLATO will use asteroseismic data and can provide stellar mass uncertainties as low as 1--3\% \citep{Rauer2014}. In consequence, with PLATO, the best uncertainties on planet radii are 2--5\% and 5--10\% on planet mass using follow-up RV measurement campaigns, which implies uncertainty in mean density between 8--18\%. Currently,  typical well-characterised planets have uncertainties of 5--10\% in radius and 10--30\% in mass. Note, that only a small fraction of planets have both characterised radii and masses, about 3\% \citep{jontof2019compositional}, which is due to the fact that transit measurements are generally independent from RV and TTV follow-up measurements. The planets with the best constrained parameters (including the rocky planet population) can be found in the PlanetS \footnote{\url{https://dace.unige.ch/exoplanets/}} catalog, recently updated by \citet{2024A&A...688A..59P}.

Along with mass and radius also temperature can be estimated by using the orbital period and stellar luminosity. 
Due to observational biases in the transit and RV measurements, which favor the detection of planets on close-in orbits, most of the small planets detected so far are warm or hot worlds, with typical blackbody temperatures ranging between 500 and 2000 K (Figure \ref{fig:temperature-density}). By comparison, the melting temperatures of rocks lie between 1400 and 1900 K at 1 bar \citep[see e.g.,][]{takahashi1986MeltingDry, katz2003NewParameterization}. It is important to note that any atmosphere surrounding a planet acts as a thermal blanket, potentially elevating surface temperatures above 2000 K. As a result, many small exoplanets may be lava worlds, with the surface dominated by a magma ocean rather than a solid mantle.

A few famous examples of temperate Earth-sized planets exist. The TRAPPIST-1 system, for example, harbours seven worlds around an ultracool red dwarf with equilibrium temperatures ranging from 400 K down to 170 K \citep{Gillon2016,Gillon2017}. Thanks to intensive campaigns of transit observations and the resonant nature of the system (constraining transit timing variations), both the radii and the masses of the planets are known to a very good precision \citep{agol_refining_2021}.

With the James Webb Space Telescope \citep[JWST,][]{Gardner2006}, we are probing the upper atmospheres of exoplanets to higher precision. However, these atmospheric observations remain very challenging for rocky exoplanets, mainly because the signals from small planets are weak, while both stellar and instrumental noise are comparable in magnitudes. The activity of the stars complicates obtaining good spectroscopic data \citep{Lim2023} but solutions might exist \citep{Rathcke2025}.  
As JWST works in the infrared, hot worlds \citep[e.g., 55 Cnc e;][]{demory2011} and temperate worlds around small stars \citep[e.g., TRAPPIST-1 b, c, and d,][]{Greene2023,Lim2023,Zieba2023,Ducrot2024} are accessible for characterization. 
Note that in \citet{Greene2023}, \citet{Zieba2023} and \citet{Ducrot2024}, the technique used to assess the presence of an atmosphere was not relying on spectroscopy, but measuring the thermal emission during a secondary eclipse (when the planet passes behind the star). 
In principle, with JWST it is feasible, but challenging for rocky planets, to identify an existence of an atmosphere, and detect volatile species like \ce{CO2}, \ce{H2O}, \ce{CH4}, \ce{CO}. The upcoming ARIEL (Atmospheric Remote-sensing Infrared Exoplanet Large-survey) mission \citep{tinetti2016ScienceARIEL, beaulieu2018ARIELSpace}, scheduled to launch in 2029, will be limited to larger planets as its mirror is 3 times smaller than for JWST. Accordingly, for any in-depth characterisation of temperate Earth-sized planets, only future mission concepts like the Habitable Worlds Observatory \citep[HWO,][]{natlacademy2023PathwaysDiscovery} and the Large Interferometer For Exoplanets \citep[LIFE,][]{quanz2022LargeInterferometer} bring the necessary abilities to characterise temperate worlds.

\section{The diversity of observed exoplanets}
\begin{figure}
    \centering
    \includegraphics[width=0.98\linewidth]{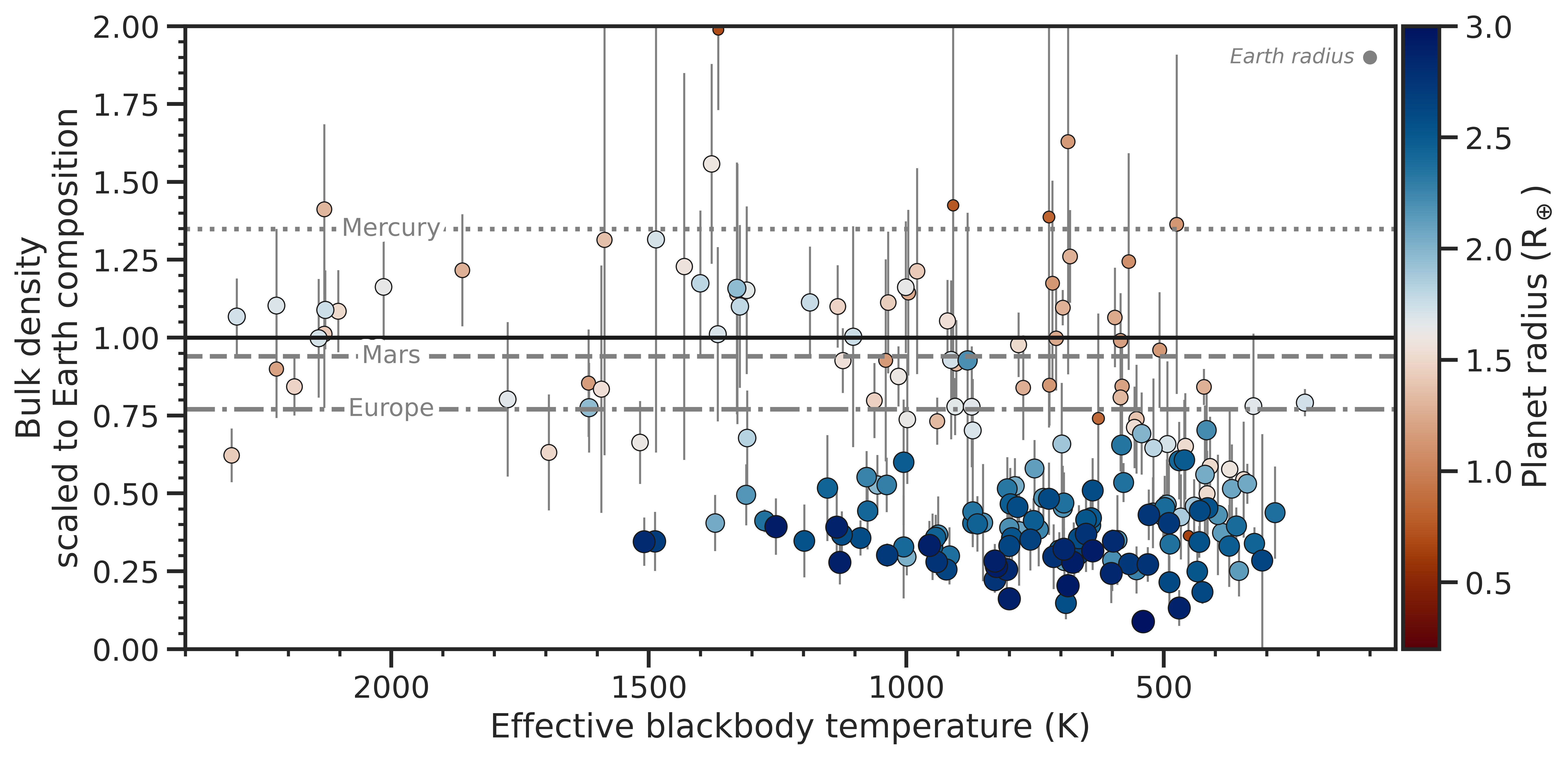}
    \caption{Confirmed exoplanets with measured radius $<3\,R_\oplus$ and measured mass $<19\,M_\oplus$, excluding planets with mass uncertainty $>50\%$. Data are from the NASA Exoplanet Archive, accessed 05/12/2024, using the most recent entries for each planet. Effective blackbody temperature is calculated assuming 30\% albedo. The $y$ axis shows bulk density normalised to the bulk density of an Earth-like composition at that planet mass, which is calculated using the scaling $R_p/R_\oplus = \left(M_p/M_\oplus\right)^{0.282}$ \citep{zeng_growth_2019, noack_parameterisations_2020}. Marker size is proportional to planet radius. Bluer (redder) colours indicate radii further above (below) 1.6\,$R_\oplus$, which is an illustrative radius often taken to separate rocky planets from volatile-rich ones \citep{rogers_most_2015}. Nonetheless, the scatter of blue markers denser than the Earth composition line and red markers less dense than this line shows that this 1.6\,$R_\oplus$ `boundary' is not hard.}
    \label{fig:temperature-density}
\end{figure}

\begin{figure}
    \centering
    \includegraphics[width=0.98\linewidth]{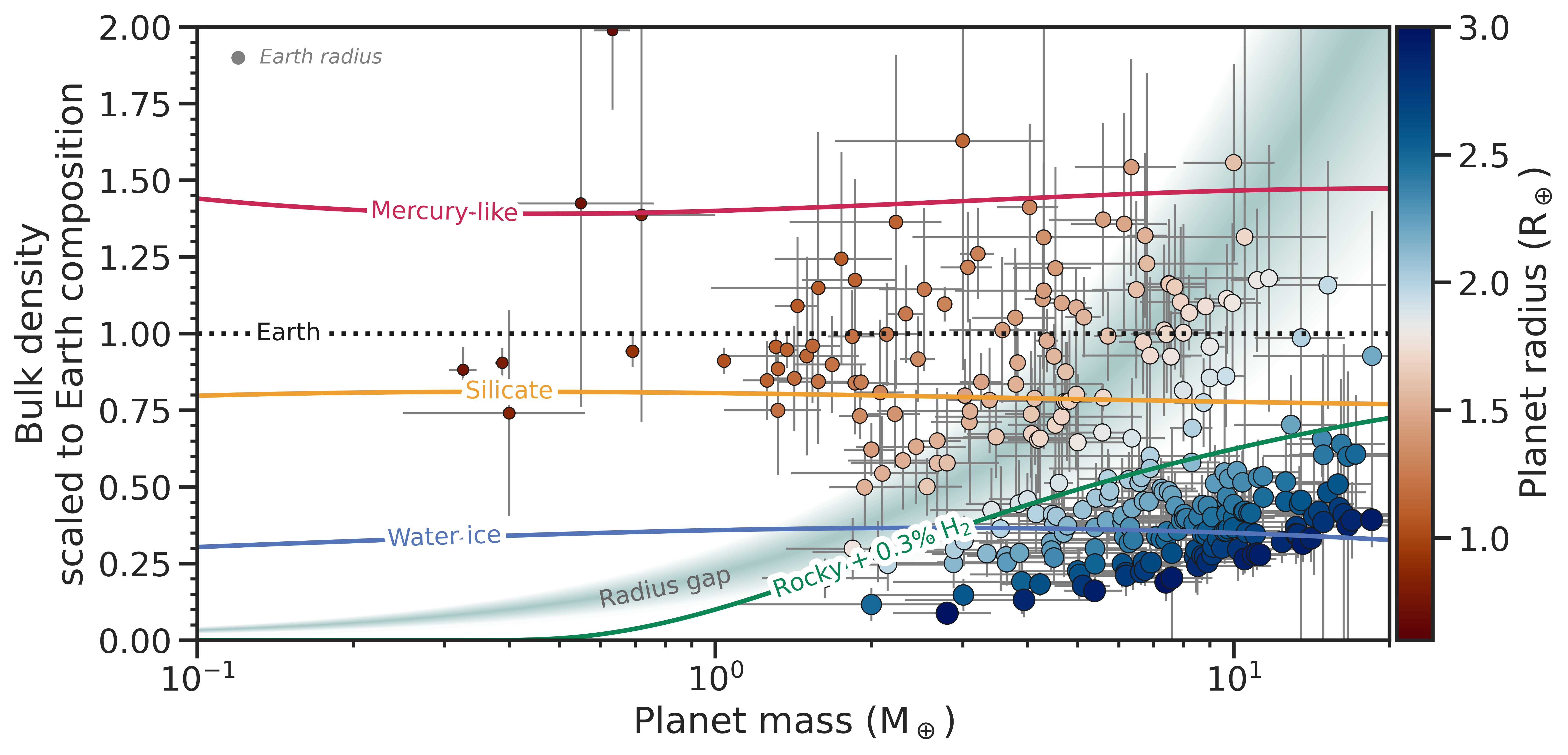}
    \caption{The same as Figure \ref{fig:temperature-density}, but plotted in terms of planet mass. The coloured lines represent the density of a planet with constant composition: Mercury-like with a 70\% iron core by mass (red, \citealt{baumeister2023ExoMDNRapid}); a theoretical pure-silicate body (yellow, \citealt{baumeister2023ExoMDNRapid}); a theoretical pure-water ice body (blue, \citealt{zeng_growth_2019}); and a theoretical Earth-like composition with 0.3\% \ce{H2} by mass at \qty{500}{K} (green, \citealt{zeng_growth_2019}). The shaded area marks the location of the radius gap, assuming a nominal range between 1.5 and 2~\RE \citep{fulton_californiakepler_2017}.}
    \label{fig:mass-density}
\end{figure}

Within the population of small exoplanets, observed bulk densities are diverse, though largely varying around Earth-like values (Figure \ref{fig:temperature-density} and \ref{fig:mass-density}). These planets are categorised based on their bulk density as super-Earths or super-Mercuries, although the majority are likely more analogous to Venus because of their high stellar irradiation, making super-Venuses a more appropriate addition to the naming convention. There are several ways to explain density variations compared to Earth; i.e., higher amounts of volatiles (water), or higher/lower core mass fractions \citep[see also the review by][]{jontof2019compositional}. In radius, the population is limited in practice by the so-called radius gap at 1.5--2.0 \RE{} (Section \ref{sec:radius-gap}).

\subsection{Radius gap}\label{sec:radius-gap}
\begin{figure}[htb!]
    \centering
    \includegraphics[width=0.9\linewidth]{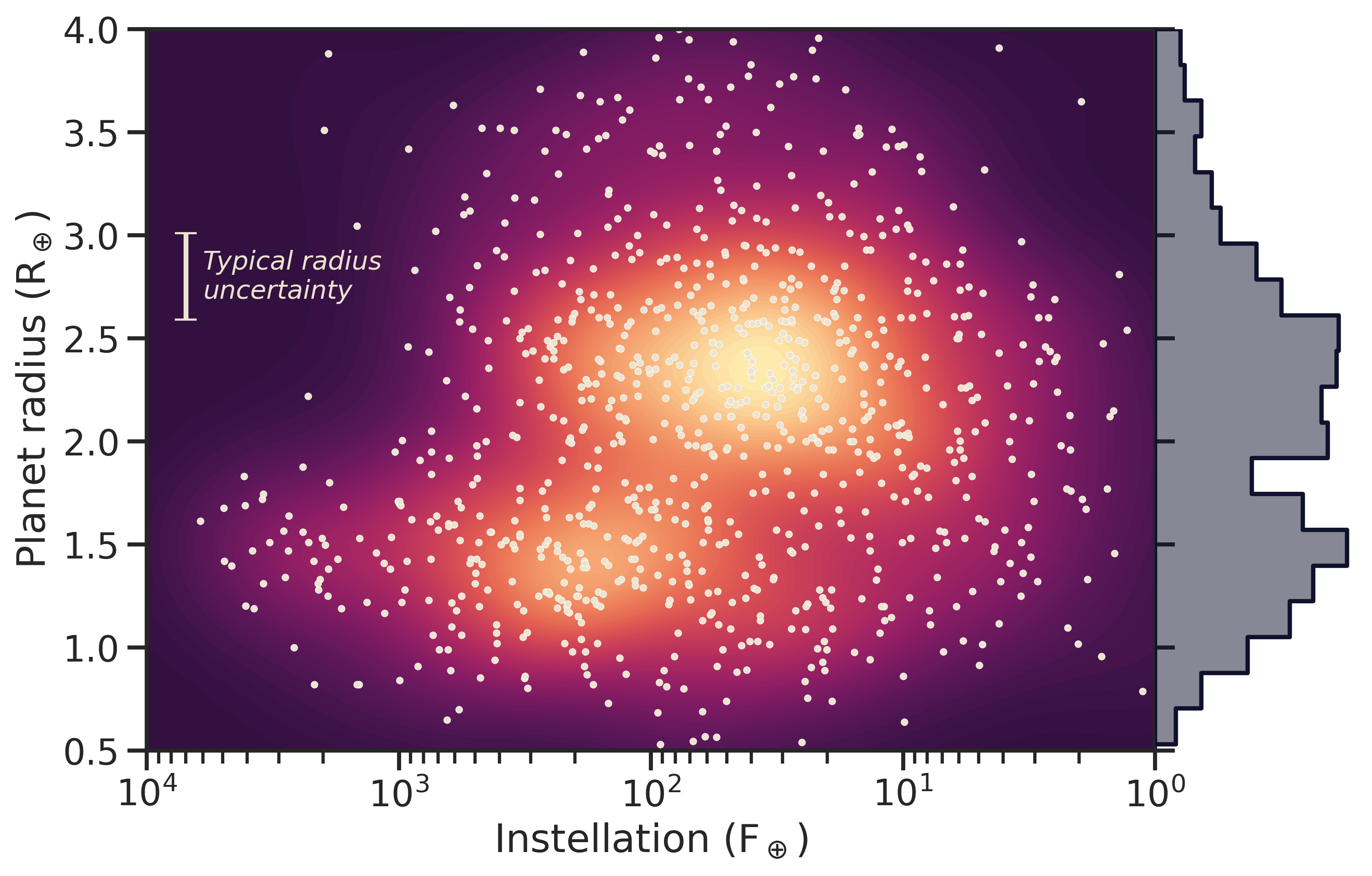}
    \caption{Distribution of observed planets with orbital periods below 100 days as a function of received stellar flux (relative to Earth) and planet radius, showing two distinct populations of planets with a scarcity of planets between 1.5 and 2~\RE. The background shows the corresponding density distribution (arbitrary units) using a kernel density estimation algorithm. The histogram on the right side shows the marginal density distribution of planet radii. Data are from the NASA Exoplanet Archive, accessed Oct 29, 2025, using the most recent entries for each planet. Only planets with a relative radius error below 10\% are shown here.}
    \label{fig:radius_gap}
\end{figure}

    The analysis of close-in planet radii with orbital periods within 100 days reveals a bimodal distribution, with two peaks at approximately \num{1.3} and \qty{2.4}{R_\oplus} \citep{owen2013KeplerPlanets, fulton_californiakepler_2017, owen2017EvaporationValley}, which has been termed the ``radius gap'' or ``radius valley'' (Fig. \ref{fig:radius_gap}). The radius gap is among the most important constraints on the interior compositions of super-Earths and the more low-density sub-Neptunes. The gap suggests that the population of super-Earths and sub-Neptunes is shaped by atmospheric loss to space of primordial gas envelopes \citep{owen2024mapping}. This idea implies that a significant fraction of observed super-Earths started their evolution with a hydrogen-dominated envelope, contemporaneous with the protoplanetary disk \citep{owen2019atmospheric, burn2024RadiusValley}. Although these envelopes are quickly lost to space after the disk lifetime of few Myrs, we expect there to be lasting chemical imprints of any hydrogen-dominated envelope in contact with an early magma ocean. Endogenic production of water, up to $\sim$1 wt.\%, is among the main consequences \citep{rogers2024fleeting}. An early hydrogen envelope has been proposed even for Earth \citep{young2023earth}, as it can fit fundamental features of the Earth (e.g., `bulk redox state',  core density) using a model of redox reactions between this hydrogen and a magma ocean.

    \subsection{Super-Mercuries}\label{sec:super-mercuries}

\begin{figure}[t]
    \centering  
    \includegraphics[width=1\linewidth]{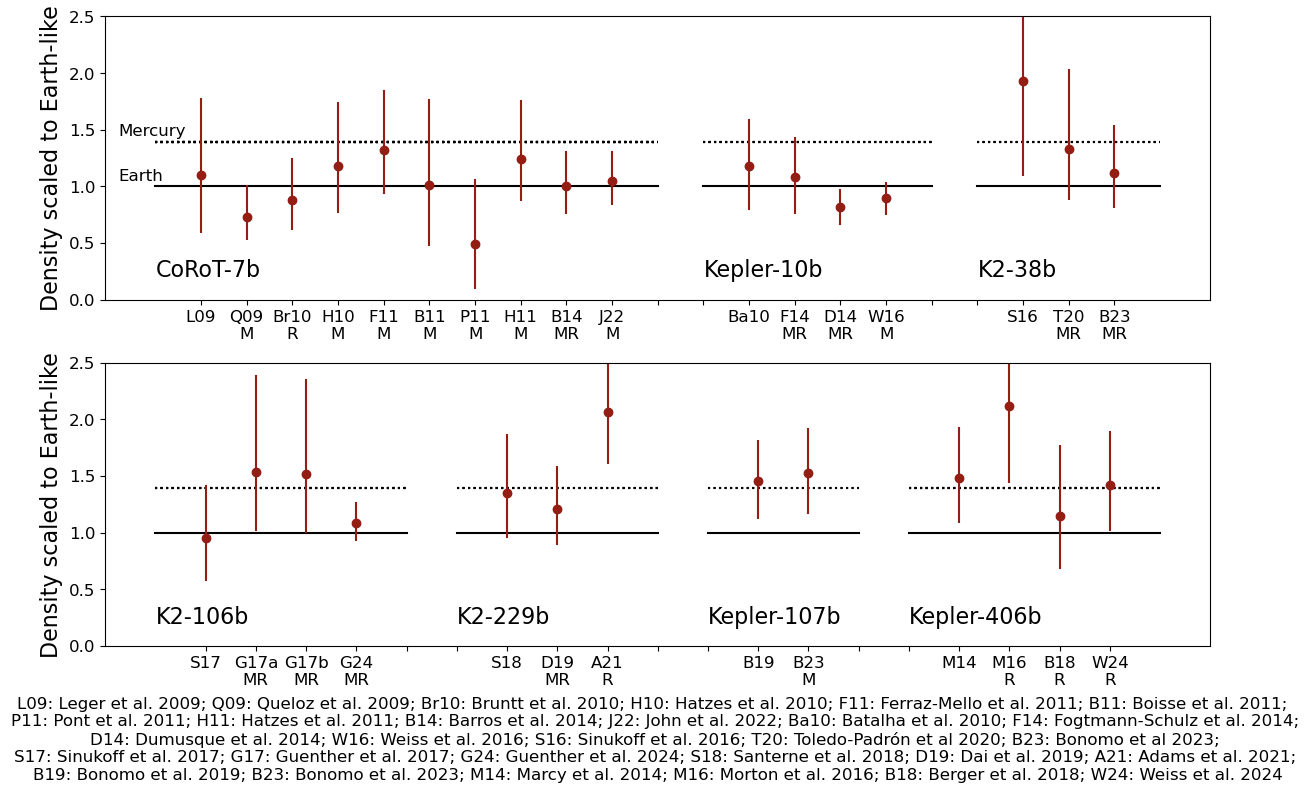}    
    \caption{Measured densities of seven different planets classified at least at some point in time as super-Mercuries in the literature, scaled to an Earth-like density and compared to a Mercury-like density (dotted lines). Labels on the x-axis refer to the respective publications; 
    ``M" refers to an update in the mass measurement and ``R" for an update in the radius measurement.
    }
    \label{fig:densities}
\end{figure}   
Most low-mass planets cluster around the bulk density of an Earth-like composition (Fig. \ref{fig:temperature-density} and \ref{fig:mass-density}). However, Fig. \ref{fig:temperature-density} also shows a number of planets with bulk densities similar to that of Mercury. Given that the radii of these planets are generally larger than the radius of Mercury, this population of planets is referred to as super-Mercuries \citep{2010Marcus}. Examples of such planets are GJ 367b and K2-229b, which both have a bulk density of $\sim$ \qty{8000}{\kg\per\cubic\m} \citep{2018SanterneK2229b,2021LamGJ367}. Interior models suggest that these planets must host a substantial iron core in order to explain the high density \citep[e.g.,][]{Wagner2012interiorstructure,2024Murgaswolf}. By comparing the bulk elemental abundance ratios of rocky planets to those of their host stars, it was shown that super-Mercuries orbit high metallicity stars \citep{LiuNi2023correlation}. 
It should be noted that the measurement errors can be large for both masses and radii, and the first planets assumed to be super-Mercuries have been revised to have lower bulk densities, owing to the ever growing observational baseline, general improvements in instrumentation, and improved determination of stellar parameters \citep[e.g., due to the GAIA catalogue,][]{gaiacollaboration2021GaiaEarly}. Figure \ref{fig:densities} shows seven example planets with their observed density over time scaled to an Earth-like density, as well as a Mercury-like density for comparison \citep[using scaling relationships from][]{noack2016water}. The first four planets (CoRoT-7b, Kepler-10b, K2-38b and K2-106b) are no longer believed to be super-Mercuries based on their revised planetary parameters. K2-229b and Kepler-406b, on the other hand, have both been first revised from Mercury-like to Earth-like in composition, but newer measurements of their radii now suggest much denser compositions. Note that this can change again in the future as their masses have not been revised for several years. Similarly Kepler-107b only has two measurements of the mass and one measurement of the radius, so the planetary parameters may still be revised in the near future. 
The current sample of rocky exoplanets still includes several super-Mercuries with reproducible, accurate density measurements.
However, \citet{Brinkman2024reanalysis} highlight that uncertainty on the radius of the host star significantly influences the measured bulk density of small planets. As a matter of fact, using updated stellar properties, they find lower bulk densities for known super-Mercury candidates. In a similar vein, \citet{Rodriguey2023Reanlysis} reanalysed the observational data for K2-106 b to find a core mass fraction of $\sim 44^{+12}_{-15} \%$ (assuming an iron-free mantle), consistent with the core mass of Earth. Accordingly, Super-Mercuries might be less frequent than previously thought. 

Multiple formation pathways can lead to the existence of super-Mercuries in theory. In one scenario, the planets form already iron-rich in a metal-enriched inner protoplanetary disk \citep{2013Wurm,johansen_nucleation_2022,mah_forming_2023}. Alternatively, the planets may have initially formed with a core mass fraction similar to Earth, but lost parts of their mantle through processes such as mantle photoevaporation \citep{1985Cameron} or giant impacts \citep{1988Benz,2022Reinhardt,Dou2024superMercuryForm}. Both the iron-rich formation theory and the photoevaporation theory predict that super-Mercuries should be the innermost planet in the system. Consequently, systems such as Kepler-107, where the iron-rich planet is not the innermost planet, are consistent with formation by giant impacts \citep{2019Bonomo107, cambioni2025CanMetalrich}. However, \citet{cambioni2025CanMetalrich} also find that the rates of mantle-stripping giant impacts is likely not sufficient to explain the sizes and abundance of observed super-Mercuries.

\section{Interior modelling of rocky planets} 

\label{sec:interior_modeling}
Interior models are an integral part in the characterization of exoplanets, providing the connecting link between observations and geophysical interpretations. To provide insights into planets' interior compositions and structures, properties of observed exoplanets (e.g., mass and radius) are compared to the same properties calculated for synthetic exoplanets. Such models consider the planets as spherical symmetric and in hydrostatic equilibrium. Differential equations to represent the profiles of mass, radius, density and temperature are solved using the physiochemical properties of chosen mineralogical phases \citep{duffy_207_2015}. The latter may be a product of thermodynamic modelling starting from a bulk elemental composition, or instead chosen \textit{a priori}. Models often consist of planets with an arbitrarily layered structure and interior mineralogy which mimic solar system planets. A considerable number of interior codes have been developed over the years \citep[e.g.,][see also subsection \ref{ssec:tools}]{sotin_mass_2007, seager2007MassRadiusRelationships, valencia_detailed_2007, rogers_framework_2010, wagner2011InteriorStructure, dorn_can_2015, dorn_generalized_2017, hinkel_star_2018, unterborn2019PressureTemperature, baumeister2020MachinelearningInference, huang2022MAGRATHEAOpensource}. 

Challenges in modelling exoplanets' interiors remain due to an inherent degeneracy between composition and bulk density. Planets with very different internal structures and mineralogies can be represented by the same calculated mass and radius, hampering a unique interpretation of the observations \citep[see section \ref{sec:non-uniqueness}]{valencia_detailed_2007, zeng_computational_2008, rogers_framework_2010}. Using other constraints --- for example, stellar composition (section \ref{sec:stellar-abundances}) --- can help to reduce the degeneracies \citep{dorn_can_2015, dorn_generalized_2017, dorn_bayesian_2017}, but a unique solution will likely never be attainable (section \ref{sec:non-uniqueness}). Understanding planetary interior structure and composition is even difficult for the Earth, where we have indirect evidence from seismology, and for solar system planets, where space missions can provide additional data such as Love numbers \citep[e.g.,][and section \ref{sec:non-uniqueness}]{park_ios_2025a}. 

In any case, to take strides towards a more accurate interpretation of planetary and exoplanetary interiors we should not forget the role of experimental studies. A wide array of high pressure and temperature experiments as well as analytical techniques have been developed to investigate and constrain the properties of geomaterials, the building blocks of any planet. Their characteristics control all the aspects of a planet, from how it is structured, to its possible large scale properties (e.g., the presence of an atmosphere or a magnetic field) and are a vital component of all the models used in the field, from mass-radius models to dynamic and thermal evolution models. In the next section we will review how equations of state are used with mass-radius models as an example for the role of material properties in the interpretation of observations. In section \ref{sec:eos-uncertainties}, we discuss the uncertainties on experimental equation of state measurements and how they affect models.

\subsection{Equations of state and high-pressure physics}\label{sec:eos}

Rocky planets are made of condensed matter (minerals and/or Fe-rich metals). As such, it is necessary to consider the bonding environment of atoms and the internal energy associated with atom-atom interaction. Accordingly, at the heart of every planet model sit equations of state (EoS) and thermal models, which encapsulate the physics of mineralogical phases under planetary interior conditions. They describe how the volume $V$ (or equivalently the density) of a material varies with pressure and temperature. In a solid, the amount of compression with pressure $P$ and temperature $T$ is characterised by the (isothermal) bulk modulus $K_T = -V \partial P / \partial V$ and its first and second derivatives $K_T' = -\partial K_T / \partial P$ and $K_T'' = -\partial^2 K_T / \partial P^2$.

EoS describe the change in volume with pressure; they are polynomials, and in certain cases their complexity can increase depending on where the polynomial is truncated. The main terms are the zero-pressure volume $(V_0)$ and the bulk modulus $(K_0)$ in the simplest polynomial. The first and second derivatives of the bulk modulus with pressure (i.e., $K_0'$ and $K_0''$) can be added if the simple formulation does not accurately represent the dataset.
The simplest way to account for temperature-induced changes in the volume is to apply a thermal expansion term to a previously determined ambient-temperature EoS. In reality, as pressure and temperature have a simultaneous effect of shrinking and inflating the unit cell volume, it is more accurate to account for both the effects in the parameterization of volume variations, and hence use a $P$-$V$-$T$ EoS that includes also a thermal model (e.g., Mie-Grüneisen-Debye, or the thermal pressure). Over the years, many formalisms have been defined to parameterize equations of state, thermal expansion, and $P$-$V$-$T$ EoS. While a precise description of all of the models is beyond the scope of this paper, additional details can be found in \citet{angel2000EquationsState}, \citet{kroll_volume_2012}, \citet{poirier_introduction_2003}, and \citet{angel_40_2018}.

Rocky planets are made up of a wide range of mineral phases, each with their own specific EoS parameters. In such cases, the equations of state of the different mineralogical phases need to be combined, which can be done in different ways. For example, with the additive-volume rule, the density $\rho_m$ of a mineral assemblage can be expressed in terms of the individual densities $\rho_i$ of every species $i$,
\begin{equation}\label{eq:additive-volume}
    \rho_m = \left(\sum_i \frac{w_i}{\rho_i}\right)^{-1},
\end{equation}
where $w_i$ is the respective weight fraction. The validity of this mixing rule has been experimentally validated at high pressures and temperatures \citep{bradley2018ExperimentalValidation}.

All EoS are based on specific thermodynamic assumptions. There is no absolute basis for defining the correct form of EoS for solids. In that regard, all EoS we use in planetary interior modelling are approximations, and depend on the data used for the parametrization. This means that, in general, there is no ``correct" choice of EoS. The validity of EoS depends on how well they fit the data; each formulation is individual because that specific set of parameters represent the best fit for the data at hand. Accordingly, the parameters from one EoS should generally not be used in another EoS formulation and it should be noted that EoS for the same materials can change and improve in time; for example, if another dataset with a higher pressure and temperature extension gets collected. In a similar manner, EoS defined for solids are generally not applicable to materials at temperatures above their melting temperature. Many of the EoS and thermal models available in literature for planetary materials are based on experiments where volume data are collected at either high pressure, high temperature, or simultaneous at both high pressure and temperature. As a consequence, the extension of the pressure and temperature range in which the materials' properties are explored strongly depends upon technological developments. 

\begin{figure}
    \centering
    \includegraphics[width=1\linewidth]{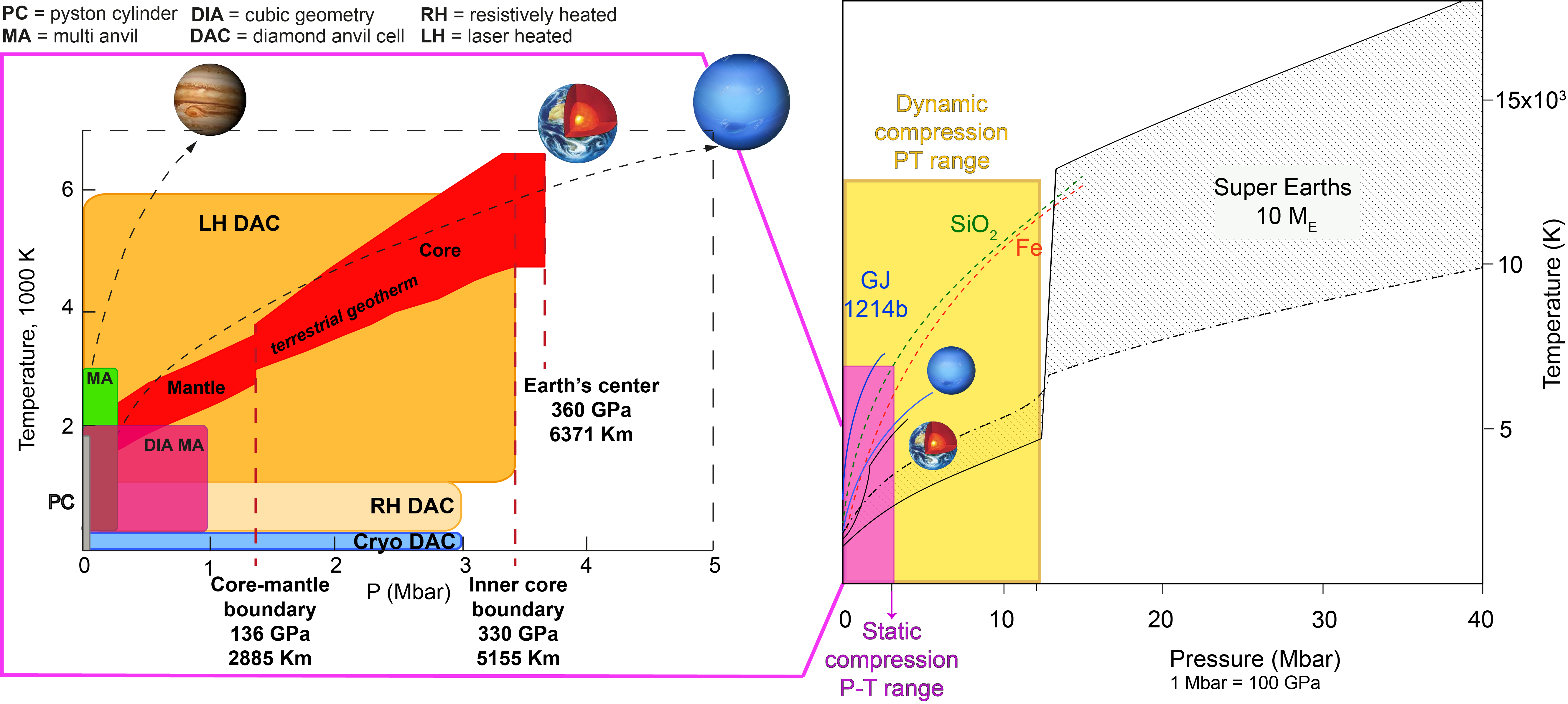}
    \caption{Pressure-temperature diagrams illustrating the ranges covered by the available experimental techniques along with some references for solar system planets and exoplanets. The figure on the right is modified from \citet{duffy_ultra-high_2019}.}
    \label{fig:PT_techniques}
\end{figure}

Experimental techniques to study EoS (i.e., density changes with pressure and temperature) have made major advances due to the improved techniques of high-pressure generation and of the measurements of density change under pressure using synchrotron X-ray (Figure \ref{fig:PT_techniques}). EoS of materials have been routinely determined experimentally down to the conditions equivalent to Earth’s deep lower mantle and core by static compression --- e.g., to $\sim$120 GPa and $\sim$2500 K for mantle minerals \citep{tange_p_2012} and $\sim$330 GPa and $\sim$3000 K for core materials \citep{tateno_structure_2010}. It is now possible to reach higher pressures and temperatures ($\sim$600 GPa, $\sim$14,000 K) with shock compression experiments \citep[e.g.,][]{wicks_b1-b2_2024,duffy_ultra-high_2019}. However, the complexity of such experiments, and the existence of few facilites where they can be performed, makes it harder to produce results for a wide set of materials. For a detailed review of high pressure techniques, see e.g., \citet{fei_high_2025}.

In principle, experimental data collected by different authors on the same material can be combined to obtain a single equation of state, as long as the datasets are consistent. This practice is commonly used by experimentalists to see how well their collected datasets compare to other. However, attention should be given to all the experimental details, especially for diamond anvil cell studies, as the experimental setup might have an effect on the measured data. More importantly, if multiple datasets are combined, the equations of state cannot be stitched together using values for the parameters from different works. It is more appropriate to combine all the data (e.g., $P$-$V$-$T$ data) and fit the desired equation of state. 

In recent years, substantial progress in computational mineral physics has also made it possible to determine EoS with numerical calculations. This techinque allows reaching large compression states, including conditions that have never been achieved in the laboratory \citep[e.g.,][]{tsuchiya2013ab}.
Ab initio methods provide a technique to numerically calculate high-pressure and -temperature properties of materials on a first-principles basis, starting from the fundamental quantum mechanical laws without introducing empirical parameters. Density Functional Theory (DFT) calculations are a widely used ab initio method to calculate the EoS of materials under extreme conditions. DFT allows the description of the structure and thermodynamic properties of materials at the atomic level, including phase diagrams, thermoelastic properties, and melting curves. DFT-derived EoS can typically not be represented by an analytical closed-form equation but are instead tabulated and interpolated between discrete data points.

Finally, upon larger compression in the \unit{TPa} regime, atom to atom interaction is no longer driven by the interaction between outer electrons. There is no distinction between outer and inner electrons; atoms lose their identity \citep{al1969electronic}. In this regime, most of the free energy of given materials is in electrons that would occupy the same energy level (the Fermi level), and these materials follow a different EoS called the Thomas-Fermi EoS \citep[e.g.,][]{poirier_introduction_2003}. Since the basic physics controlling the density in this regime differ from the physics operating at lower pressures, the mass-radius relation in this regime differs substantially from the mass-radius relation at lower pressures: close to $R\propto M^\frac{1}{3}$ at low pressures, while $R\propto M^{-\frac{1}{3}}$ at high pressures (Thomas-Fermi EoS). Most rocky planets belong to the low-pressure regime, whereas the largest giant planets would belong to the Thomas-Fermi regime.

On a side note, models of giant planets typically do not distinguish between separate silicate and iron layers, and model the ``rocky" cores as a homogeneous mixture of both \citep[e.g.,][]{hubbard1989OptimizedJupiter, nettelmann2008InitioEquation}. In fact, there may not even be a clear transition between rocky core and the surrounding volatile-rich envelope, as the respective materials may become miscible at the pressure and temperature conditions inside giant planets \citep[see e.g.,][]{vazan2022NewPerspective, young2024PhaseEquilibria, benneke2024JWSTReveals}.

\subsection{The choice of temperature profile}\label{sec:choice-temperature}

When modelling the interior of a planet, a choice has to be made on how to define the temperature profile. Temperature profiles are tightly linked to materials' thermal and transport properties (i.e., the way in which materials transfer heat). While the former (e.g., thermal expansivity) can be obtained fitting a $P$-$V$-$T$ EoS to a set of $P$-$V$-$T$ data, the latter represent one of the major sources of uncertainty because they are much harder to determine. Experiments are challenging, as they require being able to control and probe a field (i.e., electrical, magnetic, temperature) gradient. Only few groups have the capability of doing thermal conductivity experiments and to this date, not many mineralogical phases have been studied. Additionally, for some of the studied phases, there are discrepancies in the results obtained by different groups \citep[e.g., for pure Fe;][]{konopkova2016direct,ohta2016experimental,hasegawa2024inversion}. 

The interiors of rocky planets are much hotter than their surfaces, due to the large amount of heat left over from their formation, and the continuous production of heat through the decay of long-lived radionuclides. In addition, planets are very efficient at retaining this heat, in particular for planets more massive than Earth \citep{stixrude2014MeltingSuperearths}. Furthermore, iron cores are expected to be super-heated with respect to mantles because a large fraction of the gravitational energy during formation is transported to the core itself by sinking of iron-rich material. As convection in the core cannot penetrate the core-mantle boundary (CMB) layer, core cooling is further limited to conduction through the CMB \citep{stixrude2014MeltingSuperearths}.

How can we estimate the mantle and core temperatures of a planet? \citet{stixrude2014MeltingSuperearths} argues that the temperature state of rocky planets is governed by silicate melting. Cooling of a planet via melt extraction is efficient, so the top of the mantle should cool until the solidus curve is reached and cooling becomes inefficient. A similar argument can be made for the temperature at the core-mantle boundary: if the core temperature were above the solidus of the silicate mantle, a large-scale magma ocean would quickly form whose low viscosity would efficiently cool the core \citep{gaidos2010THERMODYNAMICLIMITS}. Following \citet{stixrude2014MeltingSuperearths}, the temperature $T_\text{CMB}$ at the core-mantle boundary (in K) can be estimated as

\begin{equation}
    T_\text{CMB} = 5400 (1 - \ln x_0)^{-1} \left( \frac{P_\text{CMB}}{\qty{140}{GPa}}\right)^{0.48},
\end{equation}
where $P_\text{CMB}$ is the pressure at the CMB in GPa, and $x_0$ is the mole fraction of \ce{MgSiO3} in the mantle. A value of $x_0=0.79$ produces the solidus of an Earth-like mantle composition.

Inside a planet, heat is transported through convection in the solid mantle and liquid core (for an in-depth look, see also Lourenço et al. 2025, this collection). Heat transfer in and out of a convecting parcel of material by diffusion is much slower than the typical convective time scales, making temperature changes a nearly adiabatic process. Thus, an adiabatic temperature profile is a typical assumption in interior models, with the temperature gradient over radius $r$ given as
\begin{equation}
    \frac{\mathrm{d}T}{\mathrm{d}r} = -\frac{\alpha gT}{c_{p}},
\end{equation}
where $\alpha$ is the thermal expansion coefficient, $c_{p}$ is the specific heat capacity, and $g$ is the gravitational acceleration. Note here that $\alpha$ and $c_{p}$ are material specific, and also vary with pressure and temperature. Likewise, $g$ is not constant throughout the planet \citep[although it is nearly constant in Earth's mantle, see e.g.,][]{dziewonski1981preliminary}.

Using and adiabatic temperature profile always comes with the implicit assumption that efficient convection is ongoing within the planet core and mantle. However, the rheologies, i.e., the flow behaviour of minerals,  of exoplanets with different mineralogy to Earth and at the temperature and pressure conditions of super-Earths are not well understood. Several studies have shown that the interior dynamics may be significantly different to that of Earth, for example due to the formation of a thick non-convective layer at the core-mantle boundary for planets more massive than Earth \citep{stamenkovic2012InfluencePressuredependent}, or by developing a stable mantle stratification with two convection layers \citep{spaargaren2020influence}.

Mixing length theory (MLT) provides an alternative to needing to assume an adiabatic temperature profile \citep{wagner2011InteriorStructure, Wagner2012interiorstructure}. MLT provides an estimate of the (local) convective heat flux by considering the distance a fluid parcel can move before it dissipates its thermal energy. It thus provides a way to self-consistently calculate the temperature at every point within the mantle.

Fortunately for interior modelers, the exact temperature profile plays only a minor role in controlling the bulk density of a rocky planet, and thus has only little effect on planet characterization \citep[see, e.g.,][]{seager2007MassRadiusRelationships, dorn_can_2015, unterborn_scaling_2016, thomas2016HotWater, hakim2018NewInitio}. The underlying physical reason for this is that in rocky bodies, atom-atom interactions dominate the internal energy, and thus the density depends only weakly on temperature. Instead, density in rocky planets is largely determined by pressure. This stands in contrast to gas giants, where entropy is an important part of the internal energy, and hence density is particularly sensitive to temperature. As such, temperature can have a significant effect on the inferred mass and thickness of the volatile-rich layers of a planet; for example, water oceans \citep{thomas2016HotWater} or extended atmospheres \citep[e.g.,][]{nixon2021HowDeep}.

We can estimate the first-order effect of temperature on a rocky planet's radius by assessing how the bulk density changes as a function of temperature. The temperature effect on density can be well described using the thermal expansion coefficient $\alpha$ \citep{fei1995ThermalExpansion, angel2000EquationsState}:
\begin{equation}
    \rho(T) = \rho(T_0) \exp \int_{T}^{T_0} \alpha(T) \,dT,
\end{equation}
where $\rho(T_0)$ is the density at the reference temperature $T_0$ (typically room temperature). Assuming a constant $\alpha$, this reduces to 
\begin{equation}
    \rho(T) = \rho(T_0) \exp\left[- \alpha (T - T_0)\right].
\end{equation}
Taking a characteristic value of $\alpha=\qty{2e-5}{\per\K}$ for both mantle minerals and the iron core, an \qty{1000}{K} increase in the average temperature of the planet decreases the planet bulk density by about \qty{2}{\%}, for an equivalent radius increase of only about \qty{0.7}{\%} \citep[see also][]{noack_parameterisations_2020, foley2020HeatBudget}.

However, it is to note that the temperature structure of a planet has profound consequences for its thermal evolution, the emergence of a magnetic field, and its tectonic state (see also Lourenço et al. 2025, this collection, for an in-depth exploration of the thermal evolution of planets).

\subsection{Tools for interior characterization}
\label{ssec:tools}

Mass-radius scaling laws have long formed a cornerstone of exoplanet interior modelling. Mass-radius composition curves can be well approximated by power laws in the form of
\begin{equation}
    R \propto M^\beta,
\end{equation}
with the exponent $\beta$ typically derived by fitting the power law to interior models. For rocky planets, typical values for $\beta$ are between $0.25 - 0.3$ \citep[e.g.,][]{valencia2006InternalStructure, sotin_mass_2007, wagner2011InteriorStructure, noack_parameterisations_2020}.

Over the years, these scaling laws have seen various extensions, with additional parameters such as planet bulk iron content \citep{noack_parameterisations_2020}, core mass fraction \citep{zeng2016MassRadiusRelation}, or water content and instellation \citep{turbet2020RevisedMassradius}.

As the field matured, so did the software and models available to the community, in particular in open-source codes which aim to help in the characterization of planet interiors. We aim to provide here a comprehensive overview of the ones which are of the most interest for the characterization of rocky planets and for exploring the boundaries of the population of rocky planets. Codes can be broadly categorised into four areas of application: Interior structure modelling, interior retrievals, bulk parameter estimation, and visualization. A summary of these codes with their respective links is given in Table \ref{tab:codes}.

\textbf{Interior structure codes} simulate the radial structure of a planet, such as density and temperature profiles (see Section \ref{sec:interior_modeling}).
\textit{MAGRATHEA} \citep{huang2022MAGRATHEAOpensource} is a general interior structure code written in \verb!C++! and supports up to 4 planetary layers: core, mantle, hydrosphere, and atmosphere. \textit{ExoPlex} \citep{unterborn2023NominalRanges} is a thermodynamically consistent interior structure model written specifically for predominantly rocky planets, taking into account mantle mineralogy and core chemistry, for example the fractionation of iron between mantle and core. \textit{ExoPlex} utilizes the \textit{BurnMan} toolkit \citep{myhill2023BurnManPython}, a mineral physics framework for rocky planets. \textit{BurnMan} itself allows modelling of interior structures as well. \textit{ROWS} \citep{noack_parameterisations_2020} is 1D model to calculate the internal structure and depth-dependent thermodynamic parameters for rocky planets and ocean worlds. \textit{GASTLI} \citep{acuna2024GASTLIOpensource} is an interior model which, while designed primarily for gas giants, can also model planets with extended gas and water envelopes.

\textbf{Planet composition and interior retrieval tools} aim to determine the interior composition of exoplanets based on observable bulk parameters, or try to estimate the compositional ranges, for example from the bulk stellar abundances. The interior structure codes \textit{MAGRATHEA}, \textit{ExoPlex}, and \textit{GASTLI} already include inbuilt interior retrieval capabilities. A few codes exist to estimate the distribution of individual planet interior layers: \textit{HARDCORE} \citep{suissa2018HARDCOREModel} is a parametric model to estimate the minimum and maximum core size of planets, while \textit{SMINT} \citep{piaulet2021WASP107bsDensity} provides distributions for hydrogen and water envelope mass fractions. \textit{SuperEarth.py} \citep{plotnykov2020ChemicalFingerprints} is an analytical model based on the \textit{SuperEarth} code \citep{valencia2006InternalStructure, valencia_detailed_2007} to estimate iron core mass and Fe/Si ratio of a given planet, or to estimate mass or radius given stellar Fe/Si constraints. \textit{exopie} \citep{plotnykov2024ObservationUncertainty} estimates core mass fraction of rocky planets, and water mass fraction of water worlds, along with their respective uncertainties. For full Bayesian interior retrievals, \textit{ExoMDN} \citep{baumeister2023ExoMDNRapid} and \textit{plaNETic} \citep{egger2024UnveilingInternal} are publicly available fast machine-learning based models. Going beyond mass and radius of an exoplanet, \textit{ExoInt} \citep{wang2019EnhancedConstraints} and \textit{ECCOplanets} \citep{timmermann2023RevisitingEquilibrium} are two codes to estimate the bulk planet composition from stellar abundances.

\textbf{Bulk parameter estimation tools} aim to predict unobserved bulk planet parameters, typically relying on the observed population of planets. This includes for example estimating the mass of a planet based on its radius, or vice versa. Most of these work in a probabilistic way to estimate the posterior distributions of unseen parameters, using the observed population of exoplanets as a basis, for example \textit{Forecaster} \citep{chen2017ProbabilisticForecasting}, \textit{BEM} \citep{ulmer-moll2019ExoplanetMassradius}, \textit{MRExo} \citep{kanodia2019MassRadius}, and \textit{spright} \citep{parviainen2023SprightProbabilistic}. In a way, these tools represent more sophisticated, multi-dimensional, non-parametric versions of the mass-radius relations and scaling laws. In addition, for multi-planet systems with planets both above and below the radius gap, \textit{EvapMass} \citep{owen2020TestingExoplanet} allows estimations of the minimum masses of planets above the radius gap.

Lastly, a few open-source tools exist which aid not directly in modelling of planets, but instead in the \textbf{visualization} of exoplanet data. \textit{mr-plotter} \citep{castro-gonzalez2023UnusuallyLowdensity} and \textit{MARDIGRAS} \citep{aguichine2024MardigrasVisualization} are Python-based tools to generate mass-radius plots. A similar tool, named \textit{exoplanet}, exists for Mathematica \citep{zeng2021NewPerspectives}, and has also been translated to Python \citep[\textit{pyExoRaMa},][]{francesco2022PyExoRaMaInteractive}.

\begin{table}[htb!]
    \caption{Open-source codes for modelling and characterising low-mass exoplanets.}
    {\footnotesize
    \begin{tabular}{@{}p{2.5cm}@{}p{6cm}l@{}}
        \toprule
        \textbf{Code} & \textbf{Link} & \textbf{Reference} \\ 
        \midrule
        \multicolumn{3}{l}{\textbf{Interior structure models}} \\
        MAGRATHEA & \href{https://github.com/Huang-CL/Magrathea}{github.com/Huang-CL/Magrathea} & \citet{huang2022MAGRATHEAOpensource} \\
        \multicolumn{3}{p{13cm}}{\textit{General 1D structure model supporting planets with core, mantle, hydrosphere, and atmosphere.}} \\[0.75ex]
         
        ExoPlex & \href{https://github.com/CaymanUnterborn/ExoPlex}{github.com/CaymanUnterborn/ExoPlex} & \citet{unterborn2023NominalRanges}\\
        \multicolumn{3}{p{13cm}}{\textit{Thermodynamically self-consistent interior model for rocky planets.}} \\[0.75ex]
         
        BurnMan & \href{https://github.com/geodynamics/burnman}{github.com/geodynamics/burnman} & \citet{myhill2023BurnManPython}\\
        \multicolumn{3}{p{13cm}}{\textit{Mineral physics framework for modelling planetary interiors.}} \\[0.75ex]

        GASTLI & \href{https://github.com/lorenaacuna/GASTLI}{github.com/lorenaacuna/GASTLI} & \citet{acuna2024GASTLIOpensource}\\
        \multicolumn{3}{p{13cm}}{\textit{A coupled interior-atmosphere model for mini-Neptunes and gas giants.}} \\[0.75ex]

        ROWS & \href{https://github.com/FUB-Planetary-Geodynamics/ROWS_Interior-Structure-Model}{github.com/FUB-Planetary-Geodynamics/ROWS\_Interior-Structure-Model} & \citet{noack_parameterisations_2020}\\
        \multicolumn{3}{p{13cm}}{\textit{A Python tool for modelling the internal structure of rocky planets and ocean worlds.}} \\[0.75ex]

        \midrule
        \multicolumn{3}{l}{\textbf{Planet composition and interior retrieval tools}} \\
        HARDCORE & \href{https://github.com/gsuissa/hardCORE}{github.com/gsuissa/hardCORE} & \citet{suissa2018HARDCOREModel}\\
        \multicolumn{3}{p{13cm}}{\textit{Estimates minimum and maximum core sizes of planets.}} \\[0.75ex]
                
        SMINT & \href{https://github.com/cpiaulet/smint}{github.com/cpiaulet/smint} & \citet{piaulet2021WASP107bsDensity}\\
        \multicolumn{3}{p{13cm}}{\textit{Posterior distributions for \ce{H}/\ce{He} and \ce{H2O} mass fractions.}} \\[0.75ex]
        
        SuperEarth.py & \href{https://github.com/mplotnyko/SuperEarth.py}{github.com/mplotnyko/SuperEarth.py} & \citet{plotnykov2020ChemicalFingerprints}\\
        \multicolumn{3}{p{13cm}}{\textit{An analytical model for calculating the interior parameters of exoplanets.}} \\[0.75ex]
        
        exopie & \href{https://github.com/mplotnyko/exopie}{github.com/mplotnyko/exopie} & \citet{plotnykov2024ObservationUncertainty}\\
        \multicolumn{3}{p{13cm}}{\textit{Find the interior structure error of core, water, or atmospheric mass fractions.}} \\[0.75ex]
        
        ExoMDN & \href{https://github.com/philippbaumeister/ExoMDN}{github.com/philippbaumeister/ExoMDN} & \citet{baumeister2023ExoMDNRapid} \\
        \multicolumn{3}{p{13cm}}{\textit{Fast interior retrievals with neural networks}} \\[0.75ex]
        
        plaNETic & \href{https://github.com/joannegger/plaNETic}{github.com/joannegger/plaNETic} & \citet{egger2024UnveilingInternal} \\
        \multicolumn{3}{p{13cm}}{\textit{Neural network-based Bayesian interior retrieval framework.}} \\[0.75ex]

        ExoInt & \href{https://github.com/astro-seanwhy/ExoInt}{github.com/astro-seanwhy/ExoInt} & \citet{wang2019EnhancedConstraints}\\
        \multicolumn{3}{p{13cm}}{\textit{Devolatilize stellar abundances to produce rocky bulk composition.}} \\[0.75ex]
        
        ECCOplanets & \href{https://github.com/AninaTimmermann/ECCOplanets}{github.com/AninaTimmermann/ECCOplanets} & \citet{timmermann2023RevisitingEquilibrium}\\
        \multicolumn{3}{p{13cm}}{\textit{Equilibrium condensation code to determine rocky composition from stellar abundances.}} \\[0.75ex]
        
        \midrule
        \multicolumn{3}{l}{\textbf{Bulk parameter estimation and population fitting tools}} \\
        spright & \href{https://github.com/hpparvi/spright}{github.com/hpparvi/spright} & \citet{parviainen2023SprightProbabilistic} \\
        \multicolumn{3}{p{13cm}}{\textit{Predicts planetary masses, densities, and RV semi-amplitudes from radii (or vice versa).}} \\[0.75ex]
        
        Forecaster & \href{https://github.com/chenjj2/forecaster}{github.com/chenjj2/forecaster} & \citet{chen2017ProbabilisticForecasting} \\
        \multicolumn{3}{p{13cm}}{\textit{Estimates mass from radius (or vice versa) using probabilistic mass-radius relations.}} \\[0.75ex]
        
        BEM & \href{https://github.com/soleneulmer/bem}{github.com/soleneulmer/bem} & \citet{ulmer-moll2019ExoplanetMassradius} \\
        \multicolumn{3}{p{13cm}}{\textit{Predicts planetary radii based on bulk planetary and stellar parameters.}} \\[0.75ex]
        
        MRExo & \href{https://github.com/shbhuk/mrexo}{github.com/shbhuk/mrexo} & \citet{kanodia2019MassRadius}\\
        \multicolumn{3}{p{13cm}}{\textit{Nonparametric probabilistic framework to model planet populations using up to four observables.}} \\[0.75ex]
        
        EvapMass & \href{https://github.com/jo276/EvapMass}{github.com/jo276/EvapMass} & \citet{owen2020TestingExoplanet}\\
        \multicolumn{3}{p{13cm}}{\textit{Predicts minimum masses of mini-Neptunes in multi-planet systems.}} \\[0.75ex]

        \midrule
                
        \multicolumn{3}{l}{\textbf{Visualization tools}} \\
        mr-plotter & \href{https://github.com/castro-gzlz/mr-plotter}{github.com/castro-gzlz/mr-plotter} & \citet{castro-gonzalez2023UnusuallyLowdensity}\\
        \multicolumn{3}{p{13cm}}{\textit{Mass-radius diagram plotting tool.}} \\[0.75ex]
        
        MARDIGRAS & \href{https://github.com/an0wen/MARDIGRAS}{github.com/an0wen/MARDIGRAS} & \citet{aguichine2024MardigrasVisualization}\\
        \multicolumn{3}{p{13cm}}{\textit{Interactive mass-radius relationship visualization tool.}} \\[0.75ex]
        
        exoplanet & \href{https://github.com/astrozeng/exoplanet}{github.com/astrozeng/exoplanet} & \citet{zeng2021NewPerspectives}\\
        \multicolumn{3}{p{13cm}}{\textit{Mathematica tool to plot exoplanet data and histograms in the mass-radius diagram.}} \\[0.75ex]
        
        pyExoRaMa & \href{https://github.com/francescoa97outlook/pyExoRaMa}{github.com/francescoa97outlook/pyExoRaMa} & \citet{francesco2022PyExoRaMaInteractive}\\
        \multicolumn{3}{p{13cm}}{\textit{Interactive tool to investigate the radius-mass diagram, based on the tool by \citet{zeng2021NewPerspectives}.}} \\[0.75ex]
        \bottomrule
    \end{tabular}%
    }
    \label{tab:codes}
\end{table}


\section{The perils of planet categorisation}
\label{sec:rocky}
As the articles of this collection deal primarily with the geophysics of rocky exoplanets, it is reasonable to ask how we can actually determine if an observed planet is rocky in nature or not. A first remark to make is that bulk density alone is not a good indicator of planet composition due to the compression of planetary materials at high pressures. Earth has a bulk density of \qty{5515}{\kg\per\cubic\m}. A 10 Earth mass planet with the same relative core size and composition as Earth would have a bulk density of about \qty{8500}{\kg\per\cubic\m} \citep[here calculated with the parametrizations from][]{noack_parameterisations_2020}, which is more than that of pure iron at room temperature and pressure. Some degree of interior modelling incorporating high-pressure physics is therefore necessary to reliably determine the make-up of a planet.

The term "rocky planet" is not well defined. Should we still call a planet with a \qty{100}{km} deep water ocean rocky? Similarly, how big can a planet's atmosphere be before the planet no longer classifies as rocky? For the purposes of this review, we adopt a practical definition: a rocky planet is one consisting primarily of silicate minerals and metals, with an amount of surface water not sufficient to affect the observed mass, and a relatively thin atmosphere with respect to planet radius. Based on this definition, we exclude sub-Neptunes with rocky cores but thick \ce{H2} atmospheres. A super-Mercury (see Section \ref{sec:super-mercuries}) on the other hand with a predominantly metallic composition would be considered rocky. 

Computed mass-radius relations (or mass-density relations, see Fig. \ref{fig:mass-density}) can give us a first estimate of the nature of the planet. Planets above the line for (i.e., less dense than) a pure silicate composition must contain some amount of volatiles, be it in the form of a water/ice layer or an extended atmosphere, or incorporated into the mantle \citep[although this has only a secondary effect on mass and radius,][]{shah2021InternalWater} and core. Here lies one of the limitations of this method: since we do not know the exact composition and mineralogy of the planet, there is no single composition line which can divide between the volatile-rich and rocky planets. Furthermore, a planet sitting below the line of (i.e., denser than) pure silicate composition may yet have significant amounts of volatiles. A thick atmosphere or water layer may be masked by an iron core, which compensates for the low density of the volatiles. In terms of bulk density, which is what the mass-radius curves in effect depict, this scenario can masquerade as a pure silicate planet.

As we have seen in Section \ref{sec:choice-temperature}, temperature has only little effect on the bulk density of a rocky planet. However, surface temperature can affect the observed bulk density, even for a rocky planet, if this temperature is high enough for most of the mantle to be sustained in a molten state. This particular scenario could be relevant for very young planets of Earth-size or larger. In this scenario, sufficient water may dissolve into the magma to cause a radius inflation. Studies have predicted such a density variation up to several percent for a given bulk composition \citep{2021DornLichtenbergwater}. However, details about the solubility of water and hydrogen in magma oceans are missing, hampering a realistic assessment of the radius inflation induced by volatile intake. 

In that regard, mass-radius curves are most valuable for population-wide studies, such as trying to find compositional trends in observed planets, and less so in characterising the composition of individual planets. When applied to individual planets, we quickly run into the problem that the number of unknown parameters is larger than the constraining observables. This leads to the non-uniqueness problem (interior structure degeneracy), where many possible interior structures fit a given planet (section \ref{sec:non-uniqueness}).

In spite of this long-known mass-radius-composition degeneracy \citep[e.g.,][]{seager2007MassRadiusRelationships, valencia_detailed_2007, elkins2008coreless}, exoplanet researchers have sought to map out the boundaries in mass-radius space (or, sometimes, only one of the two) that might be used to delineate ``rocky'' planets. To this end, the curve of pure \ce{MgSiO3} has been more-or-less accepted as a hypothetical end-member composition for the least-dense rocky planet \citep[e.g.,][]{rogers_most_2015}. Even then, however, planets overabundant in CaO and \ce{Al2O3} due to extreme high-temperature condensation could in principle be less dense than \ce{MgSiO3} \citep{dorn_new_2019}. Otherwise, for the majority of exoplanets, it is difficult to conceive of formation scenarios that would create a world virtually free of iron (as metal or oxide). Considering the observed distribution of stellar Fe abundances, among other compositional variables, \citet{unterborn2023NominalRanges} suggest a slightly denser mass-radius curve as a limit to ``nominally rocky'' planets, with the scaling relationship $M/M_\oplus = 0.19 + 0.64 \left(R/R_\oplus\right)^{4.1}$. They note nevertheless that anomalously Fe-poor bulk compositions could place a rocky planet out of this bound. In summary, it remains likely that a planet less dense than \ce{MgSiO3} is \textit{not} rocky. As bulk density increases towards pure Fe, a planet's rocky nature might be said to become qualitatively likelier, yet for intermediate bulk densities, attempting such a categorisation benefits from other information (e.g., very high equilibrium temperatures prohibiting volatile retention).

In reality, many known exoplanets are not detected in both transit and radial velocity, and only one of mass or radius is measured. Considering this population, it would be convenient to have a ``rocky planet cutoff'' in terms of mass or radius alone. The radius gap (section \ref{sec:radius-gap}) temptingly points to such a cutoff in radius: somewhere between 1.5 and 2.0 $R_\oplus$, but most widely interpreted to be at 1.6 $R_\oplus$. Indeed, several years before this radius gap was pointed out \citep{fulton_californiakepler_2017}, a statistical study by \citet{rogers_most_2015} independently identified $\sim$1.6 $R_\oplus$ as the point where 50\% of planets in the \textit{Kepler} sample of the time are less dense than pure \ce{MgSiO3}. However, if needing to apply this rule-of-thumb, it is important to recognise the intention of \citet{rogers_most_2015} that this cutoff is not a \textit{hard} one. Figure \ref{fig:temperature-density} provides a simple illustration of the density scatter around 1.6 $R_\oplus$: there are planets \textless1.6 $R_\oplus$ with uncompressed density closer to Europa \citep[\textgreater 7\% ice by mass;][]{gomezcasajus_updated_2021}; there are planets \textgreater1.6 $R_\oplus$ relatively denser than Earth. A corollary is that predicting an unknown mass from a known radius --- using a scaling relationship or other statistical method --- may bring misleading results. Overall, although the radius gap may \textit{broadly} separate rocky exoplanets from volatile-rich ones at a population level, classifying an individual planet is best treated with caution.

Given the inconclusiveness of a radius-only cutoff towards placing an individual planet in a categorical box, applying a mass-only cutoff is even more challenging, if only because mass increases much faster than radius for a line of constant bulk composition. A small discrepancy in mass is associated with a very large change in radius, so even more mass-radius-curve intersections become possible. Labelling a planet as rocky based on mass alone is unlikely to be informative.

\subsection{The non-uniqueness problem and ways to ameliorate it}\label{sec:non-uniqueness}
As seen in Section \ref{sec:interior_modeling}, interior models allow us to build accurate descriptions of the interior of planets, given that we know of the basic setup of the planet: its mineralogy and bulk composition. This is a \emph{forward} problem that is straightforward to handle. However, when observing a planet, we are facing an \emph{inverse} problem: we do not have observational access to the parameters which need to be prescribed in the interior models, and the goal is to recover, from a given set of observational data, the corresponding interior/model parameters which reproduce the observations. Inverse problems are notoriously difficult to solve.

In the Solar System, a wealth of data is available to help constrain the interior of the planets. Measurements of the gravitational moments constrain the mass distribution inside the planets. Seismic measurements on Earth, the Moon, and Mars allow a determination of the size of the iron core. Chemical analysis of surface material and measurements of the surface heat flow, for example from the InSight lander on Mars or the Apollo missions, allow further inferences on the material properties and interior conditions of the terrestrial planets. Even despite all this, we still do not have a clear view of the interior structure and chemistry of many Solar System bodies. For example, while we can make some inferences on the core size of Venus based on its similarities to Earth, there is still no definitive answer for the size of Venus' core.

It is an even more difficult situation with exoplanets. The measurements we are able to make of Solar System bodies are largely unavailable. As discussed throughout, exoplanet measurements are mostly limited to the basic bulk planetary parameters such as mass, radius, and, consequently, bulk density, in addition all carrying potentially substantial measurement uncertainty. As a result, exoplanet interior models are inherently non-unique, because the large number of unknown parameters outweigh the limited amount of observables. One set of observable parameters can correspond to a multitude of possible planet interior scenarios \citep[e.g.,][]{valencia_detailed_2007, zeng_computational_2008, rogers_framework_2010, dorn_can_2015, dorn_generalized_2017, baumeister2020MachinelearningInference, zhao2021MachineLearning, baumeister2023ExoMDNRapid, haldemann_biceps_2024}. The amount of inherent unknowns about an exoplanet interior is enormous: We do not know \textit{a priori} the bulk composition of the planet, if it has water on the surface, if it has an atmosphere, what the composition of the atmosphere is, what the mantle mineralogy is, how much light elements there are in the iron core, and so on. Further constraints are needed to properly reduce this non-uniqueness, and precise measurements of the mass and radius are necessary to have a chance at constraining the interior \citep[e.g.,][]{baumeister2023ExoMDNRapid, plotnykov2024ObservationUncertainty}.

In a more mathematical sense, the goal of an inverse problem is to find the conditional probability distribution $p(\mathbf{m}\mid\mathbf{o})$ (i.e., the uncertainty) of the unknown model parameters $\mathbf{m}$ (e.g., the mass of the iron core, thickness of an atmosphere) given observed parameters $\mathbf{o}$ (e.g., planet mass and radius). Bayesian inference models, using sampling algorithms such as Markov-chain Monte-Carlo \citep[e.g.,][]{rogers_framework_2010, dorn_can_2015, dorn_generalized_2017, acuna2021CharacterisationHydrospheres}, provide a way to sample $p(\mathbf{m}\mid\mathbf{o})$ and thus help to quantify the range of interior structures that fit an observed planet, while also directly including model and observational uncertainties. However, these models are typically slow and computationally expensive, as typically several hundred thousand interior structures need to be calculated to explore the parameter space of a single planet. In recent years, machine learning methods have been employed to speed up interior retrievals \citep[e.g.,][]{baumeister2020MachinelearningInference, baumeister2023ExoMDNRapid, haldemann2023ExoplanetCharacterization, egger2024UnveilingInternal}.

There is some additional information beyond mass and radius that, if invoked, can help constrain the ``rocky mass fraction'' of an exoplanet. The most commonly-used extra piece of evidence follows a working assumption that planets and their host stars have similar refractory and moderately volatile element abundances \citep[e.g.,][]{santos_constraining_2015, dorn_bayesian_2017, dorn_generalized_2017, plotnykov2024ObservationUncertainty}. If we presume such a correlation, if we have a measurement of Mg/Fe or Si/Fe in the stellar photosphere, then this measurement can be used as prior information on the bulk planet iron-to-silicate ratio (section \ref{sec:stellar-abundances}).

In this way, varying stellar elemental abundances is one axis of potential exoplanet compositional diversity. Having some idea of what bulk compositions to expect is informative in a Bayesian sense; it can serve as a reality check on choosing what materials to include in an interior structure model or parameter sweep. Whilst the details of an exoplanet's interior chemistry will always elude us to a (likely large) degree, section \ref{sec:compositional-diversity} will review current expectations of rocky planet compositional diversity and how it could be used to constrain planetary interior structure.

Lastly, a tangential but very strong indicator of the rocky nature of a hot, tidally-locked planet is the non-detection of an atmosphere from photometric phase curve observations. Phase curves reveal the hemispheric temperature contrast between the permanent dayside and nightside of the planet. A sufficiently-large temperature contrast indicates no heat redistribution, hence no atmosphere. This technique has identified LHS 3844 b \citep{kreidberg_absence_2019} and GJ 367 b \citep{zhang_gj_2024} as bare rocks---whilst the latter is dense enough to already expect a rock-iron composition, there is no mass measurement of the 1.3-$R_\oplus$ planet LHS 3844 b. However, atmospheric characterization -- or ruling out an atmosphere -- is and will only be available for a selection of a few planets, but not for the entire population of discovered planets.

\subsection{Uncertainties on mineral physics data and their effects on interior modelling}\label{sec:eos-uncertainties}

\begin{figure}
    \centering
    \includegraphics[width=0.95\linewidth]{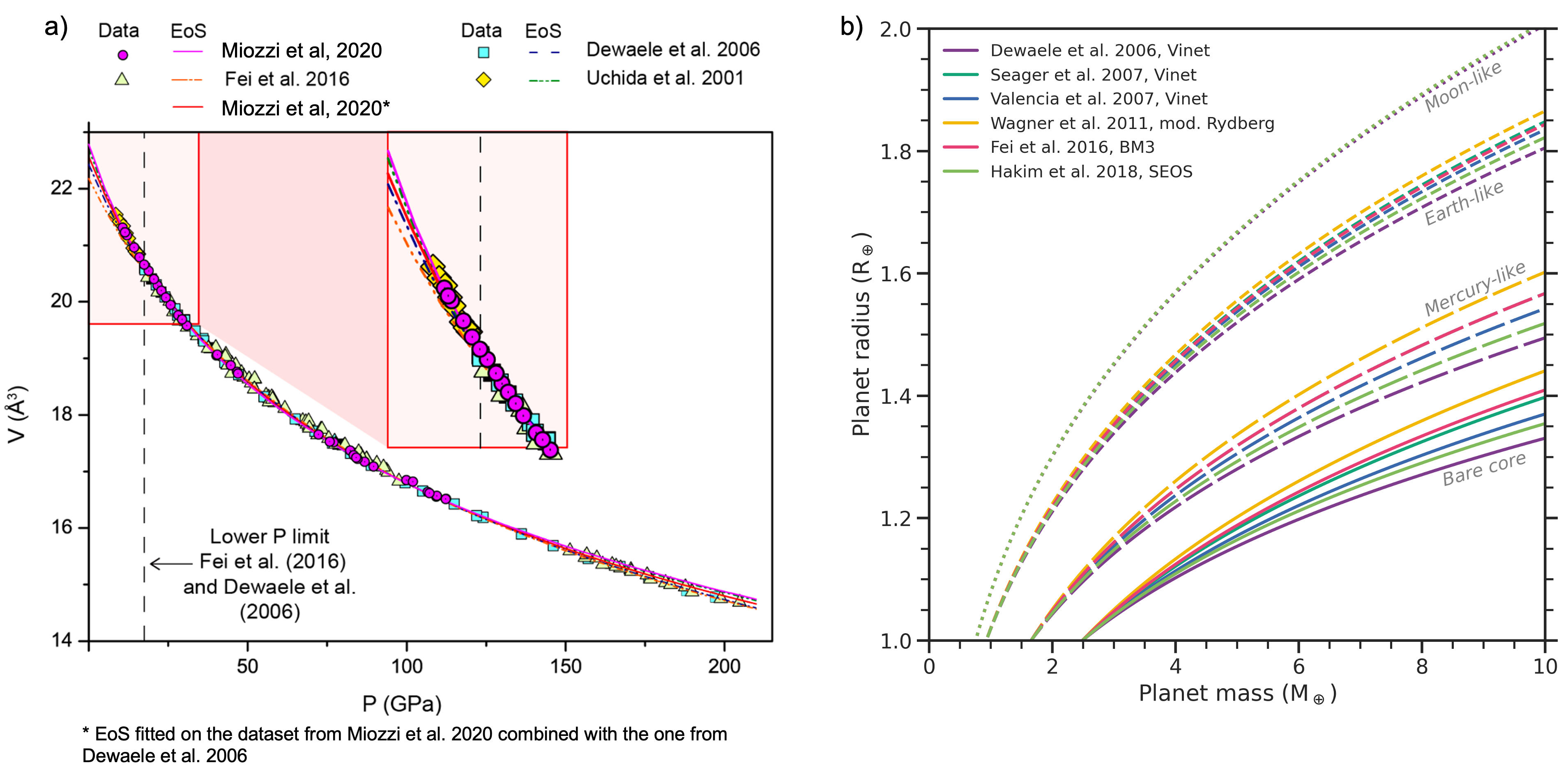}
    \caption{a) Modified from \citet{miozzi2020new}. Collection of data representing the volume variation of iron with pressure at 300K. Symbols are the experimental data, lines represent the best EoS parametrization for different authors. In the insert magnification of the lower pressure space, often challenging to investigate at high temperature. b) Variation of the calculated mass and radius, for different planets, induced by the use of a different Fe EoS, modified from \citet{hakim2018NewInitio}}
    \label{fig:EoS_Fe}
\end{figure}

As the thermodynamic parameters in the EoS are determined from the fit to experimental values, along with the uncertainty on the fit there are two more sources of uncertainties that need to be considered. One is the operator error: A $P$-$V$-$T$ equation of state fits up to six parameters with a dataset, which quite often cannot result in a unique solution when all the parameters are refined. As such, it is sometimes required to refine only a few parameters per fit, or to fix a parameter to a value obtained from the literature. Only an accurate and methodical test of all the possiblities will result in the operator chosing the best possible solution to their knowledge. This solution may be different from a refinement proposed by another author, who has chosen another formalism to best represent the data, and might be updated a few years later because a new dataset has been collected (Figure \ref{fig:EoS_Fe}a).

The second source of uncertainty is the experimental error. Data used to parametrize equations of state can come from high pressure and high temperature experiments and simulations. Depending on the techniques used for the experiments, the errors associated with the measurement of volume, pressure, and temperature during the experiment might be significant. In particular, diamond anvil cell experiments might carry a bigger uncertainty with respect to other experimental techniques (i.e. piston cylinder or multi anvil) at the experimental pressure and temperature due to the challenges in pressure calibration and the errors associated to temperature detection with spectroradiometry instead of using a thermocouple. It is the job of the researcher parametrizing the EoS to include those errors in the fit and assess how they affect the uncertainties on the parameterization. Uncertainties on DFT calculations are as well challenging to determine and they are dependent upon the specific packages that are used to calculate the intrinsic properties of matter among other \citep[e.g.,][]{mazdziarz2024uncertainty} and \citep{hakim2018NewInitio} for an example of DFT application to exoplanets and the approach to uncertainties.

Applying EoS to super-Earths often necessitates extrapolating the density beyond the pressures and temperatures range covered by the data used to parametrize the EoS. This can exacerbate the uncertainties coming from the fit of the thermodynamic parameters. In extreme cases when extrapolating to many Earth masses, EoS can assume non-physical behaviour, which is a mathematical artifact, but also an indication of the non-validity of the chosen EoS for this mass regime. \citet{unterborn2019PressureTemperature} find that the impact of uncertainties is only minor when inferring the iron core size of rocky planets. However, a larger role is played by the choice of EoS itself, as shown in \citet{hakim2018NewInitio} and Figure \ref{fig:EoS_Fe}b. In addition, uncertainties in EoS parameters such as thermal expansivity can propagate into the determination of the adiabatic temperature profile, which may increase the error in the density calculations at high pressures. Fortunately, the effect of temperature is minor for rocky planets \citep[see section \ref{sec:choice-temperature} and, e.g.,][]{hakim2018NewInitio}.

The choice of interior EoS naturally assumes a specific mineralogy, as discussed throughout. Earth and the solar system (e.g., chondrites) are often used as baseline mineralogy. Meanwhile, theoretical studies have also explored how mineralogy would change for different bulk ratios of Fe, Mg and Si \citep[e.g.,][]{spaargaren2020influence, guimond_mantle_2023, guimond_stars_2024}. However, fewer studies have been dedicated to understanding the changes in mineralogy induced by a different content of volatiles along with the major forming elements. The discovery of 55 Cancri e and its initial conjecture as a carbon-rich planet \citep{madhusudhan2012PossibleCarbonRich} prompted some investigation into carbon-rich systems. Silicon carbide (SiC), for instance, was identified as one of the first compounds to condense in a C-rich environment; several studies have investigated its high-$P-T$ behaviour \citep[e.g.,][]{wilson2014interior,nisr2017thermal}. The high abundances of sulphur and carbon in surface rocks of Mercury, indications of a very reducing oxygen fugacity (see also Section \ref{sec:fe-redox}), have prompted a wealth of studies on carbon- and sulphur-rich systems \citep[e.g.,][]{zolotov2011ChemistryMantle, namur2016SulfurSolubility, cartier2019RoleReducing, hakim2019ThermalEvolution}. Oxygen fugacity is directly related to carbon- and sulphur-rich chemistry. For example, under reducing conditions, the solubility of sulphur in silicate melts increases dramatically, so that sulphur can exist in both the silicate and metallic phases in high concentrations \citep{kilburn1997MetalSilicate, namur2016SulfurSolubility, cartier2019RoleReducing}. Carbon under reducing conditions can be stable as graphite or carbides \citep{holloway1992HighpressureFluidabsent, holloway1998GraphitemeltEquilibria}. However, the results of these studies are not often integrated in the main databases for thermodynamic modelling\footnote{e.g., \url{https://www.perplex.ethz.ch/perplex/datafiles/}}, as they did not focus on the thermo-physical properties of the stable mineralogical phases. An effort is required from the community to unveil and characterise phases and mineralogical assemblages that might become stable when both the bulk stellar composition and volatiles content is different from what observed in the solar system. It might turn out that the distribution of mineralogical phases and interior structures obtained for an Earth size planet in reality does not differ much from Earth --- or it might differ a lot, presenting the chance to study new mineralogical assemblages and phases and implement them in available thermodynamic databases, and to analyze the potential diversity of planetary interiors in a more comprehensive manner. 

\section{Exoplanet compositional diversity and constraints}\label{sec:compositional-diversity} 
\subsection{Stellar abundances}\label{sec:stellar-abundances}

It is a working hypothesis that the relative abundances of rock-forming elements are roughly preserved during planet formation in the stellar nebula. Indeed, primitive chondrite meteorites and the solar photosphere show near-identical ratios of these elements \citep{palme_solar_2014}, used to inform estimates of the bulk silicate Earth composition \citep{mcdonough_composition_1995}; the same principle may be true for extrasolar systems. Some additional theoretical support comes from models of condensation in protoplanetary disks, which also reproduce Mg/Si and Fe/Si similar between disk and planet \citep{bond_compositional_2010, carter-bond_low_2012, carter-bond_compositional_2012, moriarty_chemistry_2014, thiabaud_elemental_2015, jorge_forming_2022}.

Another, independent piece of information on planet rock-forming element composition is through elemental abundances in polluted white dwarf photospheres. The ultrahigh density and gravity of white dwarfs implies that any elements heavier than H and He in their photospheres must have likely come from a disintegrating object, possibly a planet, that fell onto it. Objects orbiting white dwarfs likely do not disintegrate onto the white dwarf all at once, and recovering the original elemental abundance and bulk composition of the original infalling object is made difficult by the uncertainty in the sinking timescales of various elements in the white dwarf's atmosphere \citep{buchan_planets_2022, brouwers_asynchronous_2023, buchan_white_2024}. Nevertheless, comparisons of elemental abundances between polluted white dwarfs and main sequence stars provide weak observational support of a planet-star compositional connection. Photospheric Ca/Fe, Mg/Fe, and Si/Fe have been observed to match in two polluted white dwarfs their main-sequence binaries, which should be chemical twins of the white dwarf progenitor stars \citep{bonsor_hoststar_2021, aguilera-gomez_host_2025}. Further, \citet{rogers_silicate_2025} observe a correlation between polluted white dwarf photospheric Mg/Si ratios and the olivine/orthopyroxene ratios of circumstellar dust, supporting the idea that the mineralogy of (former) rocky planetary material turned to dust (see section \ref{sec:mineralogy-modelling}) can be inferred from polluted white dwarf photospheric elemental abundances. \citet{trierweiler_chondritic_2023} compare the statistical distributions of various refractory and moderately-volatile abundance ratios between a sample of polluted white dwarfs and main-sequence FGKM stars, finding that the higher-metallicity stars show similar relative Fe abundances as the polluted white dwarf sample, though the spread of the latter is much wider. Testing the correlation of Fe content between that of the host star and that inferred from the planet bulk density is an active area of research (see section \ref{sec:fe-trends}).

If such a planet-star compositional correlation is real, then it would provide much-needed prior information on the interior structure of a planet. Earth is $\gtrsim$95\% Fe, Mg, Si, and/or their oxides by mass \citep{mcdonough_composition_1995, wang_elemental_2018}. By loose analogy, knowing the relative Fe/Mg/Si make-up in the bulk planet will limit the range and proportions of possible materials (e.g., in the most simplified scenario, pure Fe-metal and pure MgSiO$_3$) and thus possible mass-radius curves that fit a bulk density observation \citep[e.g.,][]{dorn_can_2015, santos_constraining_2015, dorn_generalized_2017, dorn_bayesian_2017, santos_constraining_2017, plotnykov2024ObservationUncertainty}---an approach which has been used in practice even before a potentially-rocky exoplanet was detected \citep{dubois_effect_2002, sotin_mass_2007}. In an early study of how well these informed priors on bulk planet Fe/Mg/Si translate to interior structure and composition constraints, \citet{dorn_can_2015} showed that, with some simplifying assumptions and depending on the measurement precision, using stellar abundances allows a very good constraint on the radius of a pure-Fe core, and good constraints on the mantle Fe/Si ratio. This is because elemental abundance constraints strongly correlate core size with mantle composition.

This growing body of evidence is promising; nevertheless, it remains to be seen whether compositional diversity for rocky planets within individual systems can be observationally confirmed. Even with known stellar abundances, planets forming in different disk regions, especially around K-dwarfs, may have varying core mass fractions \citep{hatalova2025compositional}, including planets that are high Ca and Al contents \citep{dorn_new_2019}. This is indicated by a combination of planet formation simulations with equilibrium condensation models of protoplanetary disks. Correlations have been reported between inferred planet bulk Fe content and other properties, namely, the age of the system \citep{weeks_link_2025} --- and indeed the orbital periods of planets in our own Solar system \citep{mcdonough_terrestrial_2021} --- indicating that multiple processes are at work to shape the bulk compositions of planets.

\begin{figure}[htb!]
\centering
\includegraphics[width=0.98\linewidth]{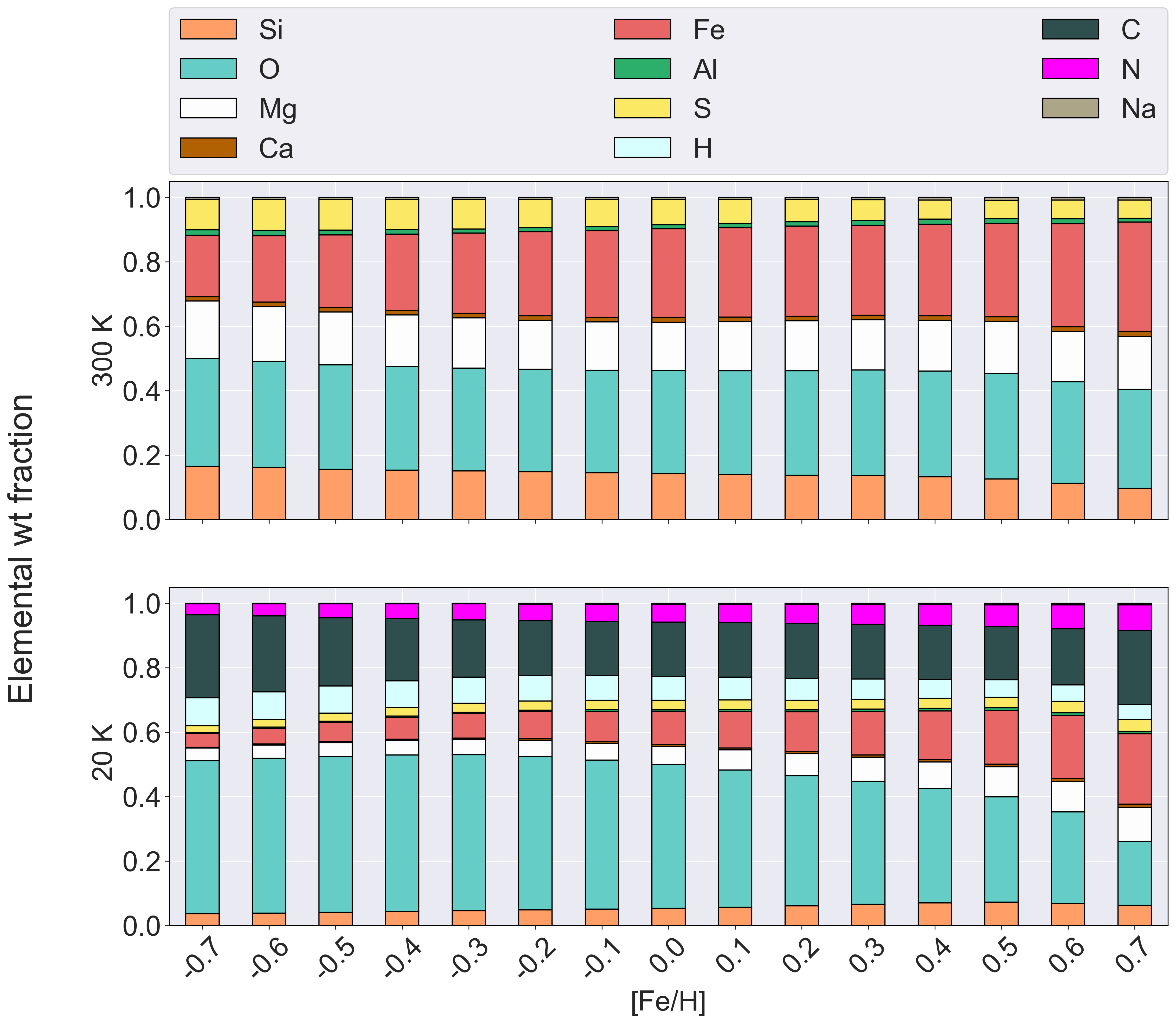}
\caption{Elemental weight fractions of planetary building blocks as a function of [Fe/H] at condensation temperatures of \qty{20}{K} and \qty{300}{K}. The data is derived from GALAH 3rd release and the Hypatia catalog.}
\label{fig:[Fe/H] trends}
\end{figure}

\citet{Bitsch2019} point out that most planet-forming simulations still rely on solar-like compositions, or on scaling the abundances of various elements using iron abundance trends. However, they show that stellar elemental abundances do not necessarily scale one-to-one with the (logarithmic) iron-to-hydrogen abundance (usually notated as [Fe/H] and normalised to the Sun). Figure \ref{fig:[Fe/H] trends} shows the elemental composition of planet-building blocks for different elements X as a function of [Fe/H] at two different condensation temperatures, based on data from the Hypatia catalog (for S, N, P) and the third release of the GALAH (Galactic Archeology with HERMES) survey (for Fe, C, Mg, O, Si, Al, Ca, Na).

\subsection{Detailed mantle mineralogy from thermodynamic modelling}\label{sec:mineralogy-modelling}

A common simplification in exoplanet interior modelling is that the mantle is a homogenous material with composition MgSiO$_3$ or (Mg,Fe)SiO$_3$. Meanwhile, xenolith samples from Earth's upper mantle and experimentally-constrained phase diagram modelling reveal a dozen different silicate mineral phases---including multiple polymorph minerals of the same chemical formula---which are thermodynamically stable depending on the local pressure and (less so) temperature. If such a detailed mineralogy were considered in interior structure models, one would require an EoS for each phase where it is stable, and combine them using the additive volume law (Eq. \ref{eq:additive-volume}). Further, although this article has somewhat focussed on the structure of the solid interior of a planet in terms of only two ``layers'', the core and mantle, silicate phase transitions will define additional layers within the mantle (e.g., the postspinel transition separating Earth's upper and lower mantle). Layered structuring can alter patterns of mantle convection, potentially drastically \citep[e.g.,][]{tackley_mantle_1995, vandenberg_massdependent_2019}, and transitions between mineral phases with differing volatile storage capacities may control how water and other volatiles are transported between interior and surface \citep[e.g.,][]{bercovici_wholemantle_2003, karato_transitionzone_2013, karato_deep_2020, guimond_mantle_2023}.

As reviewed in \citet{guimond_stars_2024}, several studies have investigated how plausible variations in bulk oxide composition (see section \ref{sec:stellar-abundances}) affect the mantle mineralogy of known and hypothetical exoplanets \citep[e.g.,][]{hinkel_star_2018, putirka_composition_2019, wang_model_2022, wang_detailed_2022, spaargaren_plausible_2023, guimond_mineralogical_2023, guimond_mantle_2023}. These studies estimate the proportions of mineral phases along a mantle geotherm, by way of Gibbs free energy minimisation routines \citep[e.g.,][]{connolly_algorithm_1987, ghiorso_pmelts_2002, stixrude_thermodynamics_2024}. A common result is that, if the observed variability of relative stellar abundances Mg/Si/Fe as well as Ca/Al \citep[e.g.,][]{hinkel_stellar_2014} indeed represents the variability of exoplanet bulk oxide compositions \ce{MgO}/\ce{SiO2}/\ce{FeO}/\ce{CaO}/\ce{Al2O3}, then the vast majority of rocky planet mantle mineralogies would fall on the spectrum of dunite to orthopyroxenite, with upper mantle ratios of olivine to orthopyroxene strongly sensitive to the Mg/Si ratio. The predicted rarity of ``exotic'' mineralogies is a direct consequence of our Sun being typical in abundance ratios compared to the solar neighbourhood \citep[see e.g.,][]{spaargaren_plausible_2023}.  

There are two important caveats to such results, both related to data paucity and already highlighted in section \ref{sec:eos} above. First, the thermodynamic data underlying Gibbs free energy minimisation models is poorly calibrated outside of Earth-like bulk compositions, and currently there is little understanding of the uncertainty due to extrapolating these compositions; e.g., to Mg/Si extremes. Second, the phase diagram of Mg-Fe-silicates at ultrahigh pressures (above several hundred GPa) is also poorly constrained, with our understanding limited mostly to molecular dynamics simulations \citep[e.g.,][]{umemoto_dissociation_2006, umemoto_phase_2017, umemoto_twostage_2011} and a handful of static-compression experiments up to $\sim$800 GPa \citep[e.g.,][]{coppari_experimental_2013, coppari_implications_2021, sakai_experimental_2016}. The unavoidably long timeframe of experimental work means that experimental constraints will often lag behind modelling developments. Knowing what exoplanet compositional diversity to expect in theory thus becomes immensely useful towards identifying which experiments should take priority; e.g., involving conditions expected on a large fraction of planets, or on particularly interesting or nearby targets.

Will variations in silicate mineralogy have a significant effect on observable masses and radii? \citet{unterborn_scaling_2016} systematically investigate the effect on mass-radius of the mantle Mg/Si, mantle Mg\# = Mg/(Mg+Fe), and presence or absence of the postspinel transition, as well as the core light element composition, showing that reasonable variations of these parameters on a $1\,R_\oplus$ planet can lead to mass discrepancies of about $0.1\,M_\oplus$, $0.4\,M_\oplus$, $0.05\,M_\oplus$, and $0.2\,M_\oplus$ respectively. That is, the Mg\# has the strongest effect on the mass-radius relationship, but its impact remains significantly smaller than the uncertainty in RV data.

Overall, although the detailed composition of an exoplanet mantle will not be constrained given observational precision in most practical applications, mineralogy has a strong influence on the subsequent geodynamic evolution of a planet, as discussed in section \ref{sec:implications}. Developing theory linking mantle compositions through to large-scale planetary consensequences is an important area of research.

\subsection{Trends and expectations in rocky planet iron contents} \label{sec:fe-trends}

Regardless of detailed phase compositions, the most important component shaping the interior structure of a rocky planet is iron, because iron is heavy, is multivalent, and where it ends up in a planet is affected strongly by gravitational and redox differentiation \citep[e.g.,][]{wordsworth_redox_2018}. The large spread in uncompressed bulk densities of the solar system terrestrial planets is attributed to an extent to the sizes of their metallic cores \citep[e.g.,][]{mcdonough_terrestrial_2021}, composed of Fe and $\sim$15--20\% alloyed elements. Between Mars and Mercury, core mass fractions (CMFs) range from $\sim$0.20--0.25 \citep{stahler_seismic_2021, khan_evidence_2023} to $\sim$0.74 \citep{margot_mercurys_2018}, posing an intriguing question as to what has caused these differences. In the solar system, Mercury is the clear anomaly, and there is no clear consensus on the origin of its large CMF: hypotheses include its mantle being stripped by early collisions \citep{cameron_strange_1988}, or the planet simply forming with more iron overall \citep[e.g.,][see also section \ref{sec:super-mercuries}]{weidenschilling_iron_1978, johansen_nucleation_2022, mah_forming_2023}. The fact that a planet's Fe content affects its bulk density so strongly is tied to Fe being the heaviest element expected in high cosmic abundance from stellar nucleosynthesis \citep{burbidge1957SynthesisElements}.

The first-order effect of iron mass fraction on bulk density means that it is somewhat tractable to define constraints on an exoplanet's iron mass fraction when thick volatile envelopes are unlikely. As discussed in the next section, bulk density is relatively insensitive to CMF specifically (except in the theoretical end-member scenario of a pure Fe core and an FeO-free mantle), but sensitive to iron mass fraction, so any inference of a CMF based on an interior structure model that presumes a pure iron core and an iron-free mantle should be interpreted as an iron mass fraction instead. In a sample of 32 exoplanets assumed to have no volatile envelopes, \citet{adibekyan_compositional_2021} estimate iron mass fractions across that sample ranging from $\sim$0 to $\sim$80\%, implying an enormous range in compositional diversity.

If rocky exoplanet refractory compositions are inherited directly from their host stars (section \ref{sec:stellar-abundances}), we would expect a 1:1 correlation between these relative number abundances (e.g., Fe/Mg, Mg/Si) as measured in stars and the same relative abundances in bulk planets. It is acknowledged that protoplanetary disk processes, including accounting for the fact that rock-forming elements are not equally refractory (``devolatilisation''), should modify these ratios \citep[e.g.,][see \citealt{guimond_stars_2024} for a recent review]{dauphas_planetary_2015, miyazaki_effects_2017, Wang2019volatility, Adibekyan2024linking}. Nevertheless, some amount of correlation between planet density and stellar iron content has been identified in the current sample, with \citet{adibekyan_compositional_2021} finding a 4:1 correlation, and more recently \citet{Brinkman2024reanalysis} finding a correlation between 1:1 and 4:1 depending on (and highlighting the importance of) the choice of linear regression method. Both studies are therefore consistent with a planet-star compositional connection. Such statistical inferences remain challenging due to high observational uncertainty and small sample sizes, however; \citet{Schulze2021probability}, for example, showed that only very discrepant planet/star iron contents could be resolved as such in the data.

\subsection{Bulk Fe redox states}\label{sec:fe-redox}

The previous section has discussed how a planet's core mass fraction is not the same as its iron mass fraction. Whatever amount of iron a planet accretes, another crucial factor affecting its core size is the interior redox (e.g., the oxygen available to react) conditions that prevailed during formation of this core.

The importance of iron chemistry on interior structure motivates searching for the underlying distribution of core mass fractions of exoplanets. If we consider only iron and oxygen, we might attempt to simplify the complicated, likely multi-stage process of core formation as the equilibrium between iron metal and iron oxide:
\begin{equation}\label{eq:iw}
\ce{ $\underset{\text{metallic melt}}{\ce{Fe^0}}$ + \frac{1}{2} O2 <=> $\underset{\text{silicate melt}}{\ce{Fe^{2+}O}}$ },
\end{equation}

where Fe and FeO are dissolved components in the metallic melt and the silicate melt, respectively. These immiscible melts would be equilibriating in the magma ocean of a young, very hot accreting planet, with gravity eventually causing the metal to settle as a planetary core. Equilibrium \eqref{eq:iw} shows that what redox state (0 or 2+) iron prefers depends on how easily an oxidising agent would be able to convert the metallic Fe into FeO: more-oxidising conditions would leave a smaller Fe core and an FeO-rich silicate mantle. This effective reactivity of \ce{O2} is quantified via the oxygen fugacity, $f_{\ce{O2}}$. For example, in the approximation from \citet{righter_redox_2012},
\begin{equation}\label{eq:diw}
\Delta{\rm IW} \approx -2 \log\left(X_{\rm Fe}/X_{\rm FeO}\right),
\end{equation}
$X$ is the molar fraction in the metal or silicate phase, here in equilibrium with each other, and $\Delta{\rm IW}$ is the dex difference in $f_{\ce{O2}}$ with respect to equilbrium \eqref{eq:iw}, known as the iron-w\"ustite or IW buffer for Fe and FeO as pure phases. In detail, $f_{\ce{O2}}$ depends on the thermodynamic activity of all involved components (in \eqref{eq:iw}, Fe and FeO) as well as pressure and temperature. Hence $f_{\ce{O2}}$ is often used as a proxy for, though does not uniquely determine, the distribution of iron between its redox states \citep{frost_introduction_1991}.

\begin{figure}[htb!]
    \centering
    \includegraphics[width=0.9\linewidth]{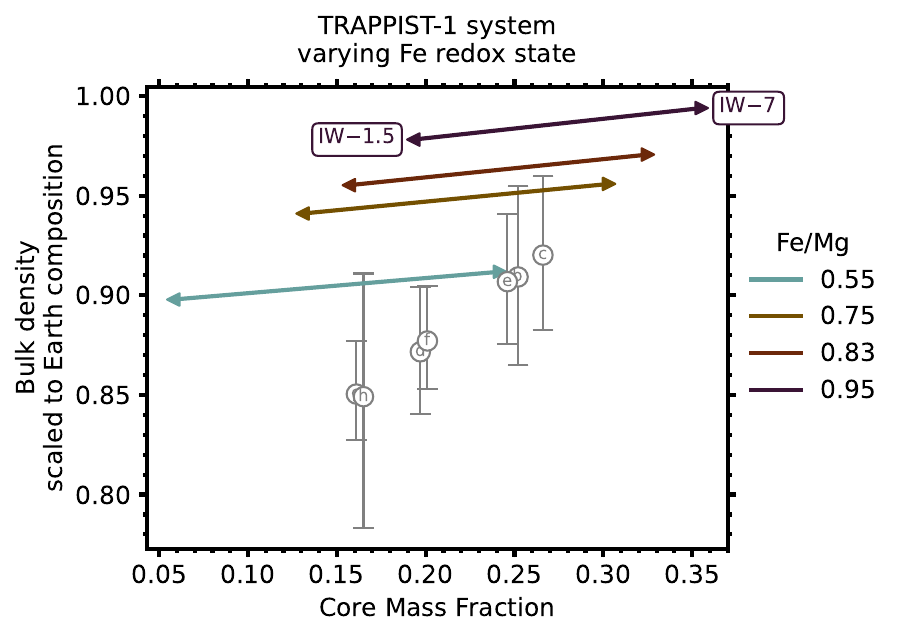}
    \caption{The effect on bulk density of partitioning iron between the core and mantle. For each line of constant bulk-planet Fe/Mg (coloured lines), the FeO mantle fraction is calculated via a given $\Delta$IW using \eqref{eq:diw}, varied from $-$7 to $-$1.5, which results in a particular core mass fraction for each $\Delta$IW and Fe/Mg. The resulting bulk density variations are much smaller than the observed bulk density uncertainties (grey errorbars) on the TRAPPIST-1 planets from \citet{agol_refining_2021}. Mantle EoS are from \citet{stixrude_thermodynamics_2024} calculated using Perple\_X and assuming an otherwise Earth-like composition. The core EoS is 100\% solid hcp iron from \citet{saxena_thermodynamics_2015}.}
    \label{fig:cmf_variation}
\end{figure}

Because planetary cores are unlikely to be pure iron, their sizes will also depend on their light element composition. Detailed core compositions are difficult to constrain even for Earth, but are informed by measurements of metal-silicate partitioning coefficients for different elements, which dictate how much of an element enters a metallic phase (so ultimately the core) or a silicate phase (so ultimately the mantle) at chemical equilibrium. These coefficients depend strongly on $f_{\ce{O2}}$ as well as temperature and pressure. At high pressures relevant for planetary core formation, many elements (but not H) become more siderophile as conditions become more oxidising \citep{suer_distribution_2023}. Core light element composition could affect core mass fraction for multiple reasons; e.g., via \textit{(i)} adding material to the core, and \textit{(ii)} decreasing the activity of metallic Fe and therefore the amount of FeO in equilibrium \eqref{eq:iw}. Hence CMF will be a complicated function of $f_{\ce{O2}}$. Moreover, the presence of impurities depresses the melting point of iron \citep[e.g.,][]{morard2017fe}, so $f_{\ce{O2}}$ at core formation can also affect for how long the core is in a liquid state (and thus the choice of EOS in an interior structure model).

Running interior structure models, however, reveals that merely re-distributing iron between core and mantle at a fixed iron budget does not have a significant effect on planet bulk density. \citet{rogers_framework_2010}, \citet{noack_parameterisations_2020} and \citet{plotnykov2024ObservationUncertainty} pointed out that increasing the iron content of a planet makes it similarly denser, regardless of whether the iron is dissolved as an oxide component in silicate solid solution, or whether it is in metallic form as a core to the planet. Figure \ref{fig:cmf_variation} shows the insignificant effect on bulk density of changing Fe/FeO. Moreover, if Fe metal droplets cannot sink in a turbulent magma ocean, a core might never form \citep{lichtenberg_redox_2021} despite the presence of this Fe. After all, iron redox chemistry in a real planet interior is much more complicated than equation \eqref{eq:iw}, and is coupled with other elements. Therefore, if we wanted to define a bulk redox state of Fe in a planet, e.g., as the total mass of Fe versus \ce{Fe^{2+}}, we could not easily extract this information from mass and radius measurements using interior structure models. Exoplanet bulk Fe redox states remain poorly constrained.

\subsubsection{Oxygen abundances from polluted white dwarfs}\label{sec:pwd}

A series of studies \citep{doyle_oxygen_2019, doyle_where_2020, doyle_new_2023} sought to measure abundances of rock-forming elements Mg, Si, Ca, Al, Fe, and O in a growing sample of PWDs, and through comparison to solar system material, assess whether these extrasolar systems had similar relative oxygen abundances. Namely, these studies aim to find $X_{\rm FeO}$ in the polluting fragments, possible for systems with enough accounted O to form oxides with all the metals Mg, Si, Ca, Al, and Fe. Assuming that the accreted material in a PWD represented the bulk mantle composition, and assuming a value for the concentration $X_{\rm Fe}$ in a metallic phase in equilibrium with this mantle composition, then an estimate of $\Delta$IW could be made using \eqref{eq:diw} \citep{righter_redox_2012}. This value of $\Delta$IW would indicate redox conditions during the parent body's differentiation, although it would not directly inform $\Delta$IW in the upper mantle or at the surface of the body insofar as the local $f_{\ce{O2}}$ is not buffered by Fe-metal here \citep[e.g.,][]{mojzsis_geoastronomy_2022, guimond_mineralogical_2023}. Another source of difficulty in learning about bulk planet redox states from $X_{\rm FeO}$ in polluting fragments is that \eqref{eq:diw} is only meaningful when FeO coexists with a separate Fe-metal phase. However, many PWDs show an excess of oxygen, such that there is enough of it to bond to all Si, Mg, Al, Ca, and Fe, with some left over \citep[e.g.,][]{doyle_oxygen_2019, brouwers_asynchronous_2023}. In this case, the application of \eqref{eq:diw} would presume that the parent body is differentiated, having formed long ago a Fe core -- the coexisting Fe-metal phase needed to define bulk redox state -- which would never be sampled in the detected `mantle' fragment anyways. Another possibility is that the body never did differentiate. Then, for fragments showing excess oxygen, a bulk redox state in terms of $\Delta$IW could not be strictly defined because there was never Fe-metal. Undifferentiated bodies may be difficult to distinguish from mantle-rich fragments in part of the PWD population \citep{buchan_white_2024}. In summary, inferring bulk redox states of parent planets using PWD oxygen abundances will require some caution, but is an intriguing avenue for population studies given enough samples \citep[see][]{buchan_white_2024}.

\subsection{Metal core composition and light element content}\label{sec:core-composition}
Metallic cores are a result of metal-silicate differentiation taking place during planetary accretion. The main driver of the process is gravity, as Fe is denser than the surrounding silicates, but it also requires the presence of a melt and consequently temperatures sufficiently high to induce large scale melting \citep[e.g.,][and references therein]{lichtenberg2022geophysical}. As discussed in section \ref{sec:fe-redox}, the Fe/FeO budget of the forming planet and the chemical exchanges potentially happening during differentiation control the size and chemistry of the forming core. Elements will partition in the metal or in the silicates depending on their chemical affinity. Light elements, such as O, C, Si, S and H, are the prime candidates to enter the metal during differentiation. Their presence can affect the melting temperature \citep[e.g.,][in the Fe-C system]{morard2017fe}, volume and density \citep{miozzi2020eutectic}, and sound velocities \citep[e.g.,][]{badro2007effect}. All of these factors are extremely relevant towards planets' large scale properties, from heat dissipation to the formation of an inner core and the driving of a geodynamo (for further discussion see Lourenço et al. 2025, this collection). 

The inventory of light elements in the core is controlled by the initial bulk composition, the reactions taking place during formation and more importantly their solubility in the metal. The latter is not pivotal to assess as it depends on the pressure and temperature of the system. Hence one light element can be more prone to enter the metal at depth, while another might exsolve from the metal when a pressure/temperature threshold is passed \citep{hirose2021light, suer_distribution_2023}. For example, metal-silicate partitioning experiments for C have shown that carbon becomes more siderophile at high temperatures but less siderophile at high pressures \citep{2013Dasguptacarbon,2014ChiCarbon,2019Malavergnecarbon}. 
Nevertheless, C is distributed roughly equally between the interior and the atmosphere of a planet \citep{suer_distribution_2023}.
Hydrogen on the other hand becomes more siderophile as the pressure increases \citep{2019Malavergnecarbon}.
The preferred incorporation of H and O in the core has also been shown by first principle simulations, making the core a potential large reservoir of water for rocky planets \citep{2020Liwater,2020YuanSteinleNeumannH}.
As a consequence, the carbon content of the core decreases with the mass of the planet, while the H content increases. 
The incorporation of N into the core depends both on the pressure and the oxygen fugacity, whereas N is only soluble in a magma ocean under very reduced conditions \citep{2003LibourelNitrogenMelt,2019GrewalNitrogenCore}. 
Even so, the largest reservoir of N remains the atmosphere \citep{suer_distribution_2023}.

The effects of light elements on the thermal equation of state for solid iron have been widely investigated \citep[e.g.,][and references therein]{hirose2021light, litasov_composition_2016}. However, less is known about their effects on liquid metals, as the experimental field is still young. 
Exoplanet metal cores are often modelled as a pure iron sphere, and one single equation of state is used, although a few studies have considered variable core light element compositions, as a free parameter \citep[see also section \ref{sec:mineralogy-modelling}]{unterborn_scaling_2016, plotnykov2024ObservationUncertainty, haldemann_biceps_2024}, or informed by solubility in metal \citep{2024Luocorewater}. Any treatment of changes of state (e.g., liquid to solid) and their associated change of EoS will be complicated by the need to couple a thermal evolution model to the hydrostatic structure equations, given that the core is expected to lose heat over time, although solid-liquid phase changes might be considered statically \citep[e.g., in][]{haldemann_biceps_2024}.

\subsection{Devolatilization trends}

Proximity to the host star plays a critical role in planetary devolatilization, the process whereby the element abundances of planets would be to a degree sorted by condensation temperature. Elements can be broadly characterised into two categories: Refractory elements, such as Al, Ca, Mg, Si, and Fe, form compounds with high condensation temperatures, typically above \qty{1300}{K}. Volatile elements, including H, S, N, C, O, and the noble gases, have low condensation temperatures below \qty{1100}{K} \citep[e.g.,][]{taylor2001SolarSystem}. While O and C are generally considered volatile, it should be noted here that they cannot be univocally classified as purely volatile or refractory, as they condense in both refractory and volatile compounds \citep{unterborn_effects_2017, wang_volatility_2019, spaargaren2025ProtoplanetaryDisk}. For example, \citet{wang_volatility_2019}, based on solar and Earth abundances, estimate that 20\% of O is incorporated into refractory silicate mineral phases \citep[see also][]{lodders2003SolarSystem}. Around carbon-enriched stars, carbon can be part of the first condensates \citep[e.g.,][]{madhusudhan2012PossibleCarbonRich}. 

Higher temperatures in the inner regions of a protoplanetary disk drive more intense volatile depletion, whereas cooler conditions at greater distances allow a larger fraction of volatiles to remain. Consequently, the radial distance and the associated thermal environment shape the observed volatility trends among planetary bodies.

Building on these general principles, \citet{Wang2019volatility} investigate the volatility of protosolar and terrestrial elemental abundances. They quantify devolatilization processes in a protoplanetary disk by comparing protosolar abundances with the composition of Earth using an improved combination of solar photospheric and CI chondrite abundances. This refined approach is crucial, as it allows for a more precise determination of the 'protosolar' composition, providing a robust baseline for comparison with terrestrial data. They derive a volatility trend, which indicates that elements with low condensation temperatures are more strongly depleted in Earth’s composition compared to the protosolar material. 

\begin{figure}[htb!]
    \centering
    \includegraphics[width=0.8\linewidth]{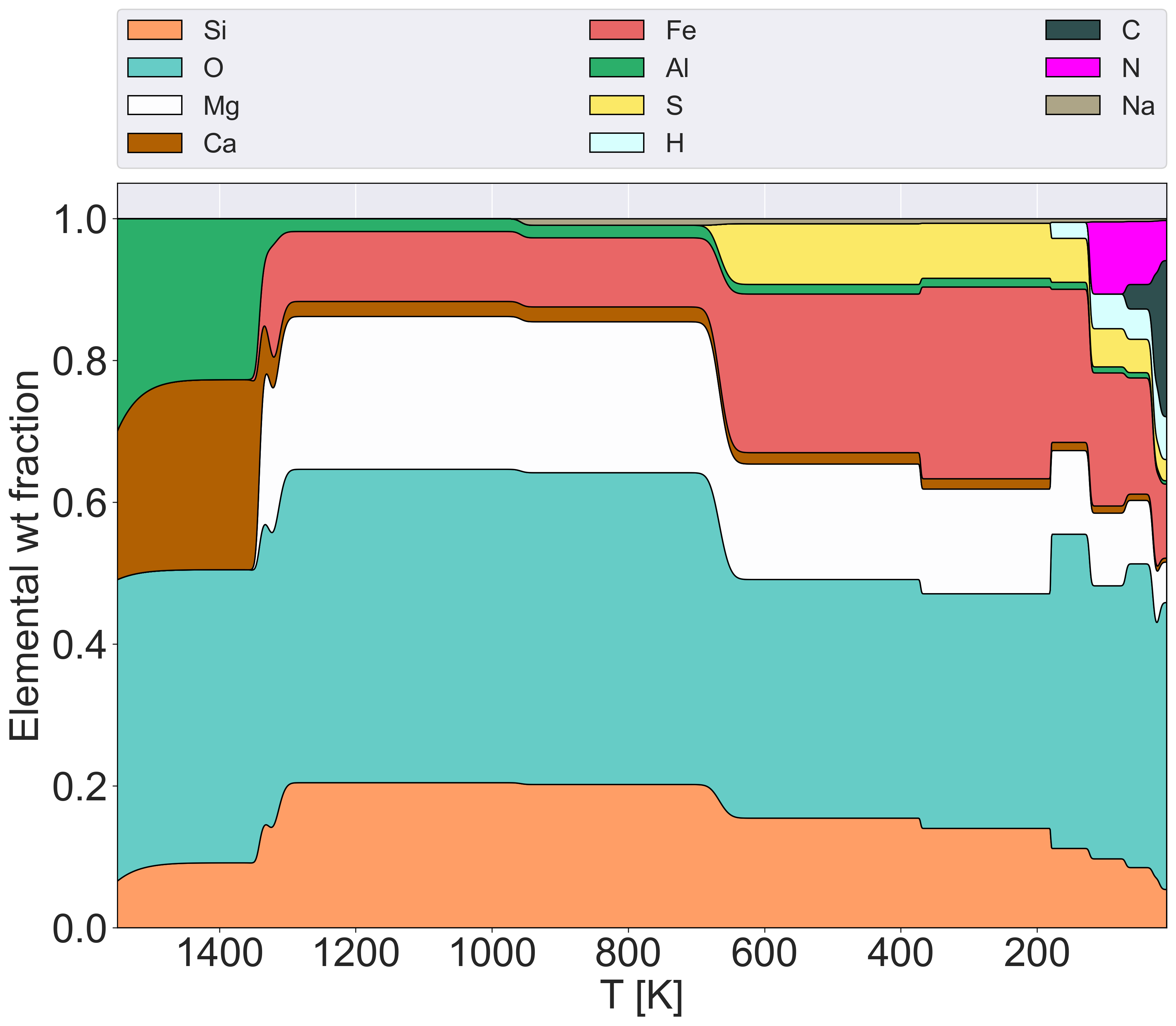}
    \caption{Devolatilization trends for different elements as a function of temperature in a protoplanetary disk, assuming a Sun-like [Fe/H]=0.}
    \label{fig:vola_trend}
\end{figure}

Outside the Solar System, it is not possible to rely on direct measurements. Consequently, complex models such as GGChem \citep{Woitke2018} and FastChem \citep{Stock2018} are used to simulate the condensation processes in a protoplanetary disk. The condensation curves shown in Figure \ref{fig:vola_trend} are based on a much simpler model by \citet{Bitsch2019}, providing a first approximation of a devolatilization trend in a protoplanetary disk around a star with [Fe/H] = 0.0 (i.e., solar; see section \ref{sec:stellar-abundances}). Here, elemental weight fractions of various condensed elements are plotted as a function of temperature within the disk. The lower the temperature--that is, the farther from the host star--the more volatile elements condense. Oxygen is an exception here, as oxygen-bearing compounds can already precipitate to form minerals at very high temperatures as discussed above.

\subsection{Volatile inventories}
\label{sec:devolatilisation}
In contrast to gas giants, volatile elements like hydrogen, carbon, oxygen, and nitrogen are only a minor component in rocky planets.
As discussed above, it is important to note that significant fractions of oxygen is in the refractory phase and therefore not counted as part of the volatile budget. Earth, the best studied rocky planet, contains on the order of $10$ ppm N, $10^2$ ppm of H, $10^3$ ppm C, and $\sim 30\,\%$ O \citep[by mass,][]{mcdonough_composition_1995, wang_elemental_2018}.
Taking into account that the oxygen abundance of the primitive mantle is estimated to be $\sim 44\,\%$, the volatile O budget of the Earth is on the order of $10^2$ ppm \citep{oneill2005mantle}.
While the exact composition of Mars is uncertain, there is both geophysical and cosmochemical evidence that Mars has a higher abundance of volatile elements than Earth \citep{2022KhanMars}.
Despite their low abundances, volatile elements play an important role in shaping the planet in terms of it geological and biological evolution \citep[e.g.,][see also Guimond et al. 2025, this collection]{2014MikhailSverjensky,2014Badrocore,2015Armstrong,2019Dehant,2023KrijtCHNOPS}.  

Constraining the bulk volatile inventory of rocky planets is difficult as these elements are distributed over the core, mantle, surface and potential atmosphere of a planet \citep[see][for a recent review]{suer_distribution_2023}, on top of the fact that devolatilization (section \ref{sec:devolatilisation}) is a poorly-known aspect of planet formation models. However, one important process determining the distribution of volatile elements in a planet is their solubility in an early magma ocean, and possible degassing from this magma ocean as it crystallises. Rapid formation, giant impacts, and radioactive decay cause rocky planets to melt and host (possibly multiple stages of) magma oceans during their early times \citep{2012ElkinsTanton, 2020Davies,2021Chao,2023Johansen}. 
Another important process is the metal-silicate partitioning in this magma ocean, during core-mantle differentiation (section \ref{sec:core-composition}).

Therefore, it can be important that interior structure models take the distribution of volatile elements into account, in particular for young or highly irradiated planets, as they have the potential to host magma oceans. Namely, the total radius of a rocky planet with a fully molten mantle and a given bulk water mass fraction can change by up to $25\,\%$ depending on whether the water is only at the surface, or is also dissolved in the molten mantle and core \citep{2021DornLichtenbergwater,2024Luocorewater}.  The interior water storage capacity of solid exoplanets, on the other hand, is much more limited \citep[see also Guimond et al. 2025, this collection]{iwamori2007transportation, guimond_mantle_2023}. Therefore, the mantle outgasses a significant fraction of its water as it solidifies \citep[e.g][and Guimond et al. 2025, this collection]{2021Gaillardmagmaatm}. 

The only direct way to detect volatile elements on a planet is to measure the composition of its atmosphere. However, there has been no clear detection of an atmosphere on a rocky exoplanet at the time of writing this article. 
Instead, the featureless spectra measured by JWST are consistent with both high mean molecular weight atmospheres composed of e.g., CH$_4$, CO$_2$ or H$_2$O, or with no atmospheres  at all \citep{2023LustigYaeger,Zieba2023,zhang_gj_2024,2025Alam}.  
The best candidates to study the volatile component of rocky planets are underdense lava worlds, where dayside temperatures are hot enough to sustain a permanent magma ocean or pond, yet the stellar irradiation is not intense enough to strip any atmosphere of volatiles completely \citep[e.g.,][]{2023Piette}. One very promising example of such a lava-world is 55 Cancri e, which potentially hosts a CO- or CO$_2$-rich atmosphere outgassed from an underlying magma ocean \citep{2024Hu55cn}.
The analysis of the atmosphere of 55 Cancri e, however, is complicated by the variability of the observed occultation depth and phase curve \citep{2024Patel55Cnc3}.

\section{Implications for exoplanet geodynamics}\label{sec:implications}

Exoplanet astronomy is, by design, focussed mostly on the observable properties of a planet: mass, radius, and orbit, but also surface temperature and composition for atmosphere-less rocky bodies, or atmospheric chemistry otherwise. The hope to identify a planet with not only habitable surface conditions, but also signs of life, is one of the main drivers of the field, such that immense investments have been made into observational capabilities both in space and on ground (see also Lagage et al. 2025, this collection). For the most efficient use of future observatories, an optimal pre-selection of observational targets are needed, which led to the growing community focussing on geoscience aspects of exoplanets. After the first detection of rocky exoplanets almost two decades ago, theoretical studies \citep{valencia2006InternalStructure,elkins2008coreless,noack2014can,dorn2018outgassing} quickly showed, that next to the planetary mass and radius, the interior structure is \textit{the} main factor influencing the geodynamics of planets, and consequently their long-term evolutionary processes.. 

While several of the first theoretical studies of super-Earths took Earth as a benchmark for a rocky planet, and then explored how planetary processes such as plate tectonics, magnetic field strength, or volcanic activity would change if a model Earth were scaled up to higher planetary masses \citep{valencia_detailed_2007,van2011plate,oneill2007conditions,noack2014plate,noack2017volcanism,kite2009geodynamics,driscoll2011optimal,bonati2021structure}, other studies showed that diverse interior structures also have geodynamic consequences \citep[e.g.,][]{elkins2008coreless,noack2014can, dorn2018outgassing,lichtenberg_redox_2021, bonati2021structure, baumeister2023RedoxState}. Incomplete differentiation of the core may both control and be controlled by interior redox conditions (see also Section \ref{sec:fe-redox}). For example, inefficient formation of the core on massive super-Earths due to physical iron-metal entrainment would lead to a reducing mantle and hence a reducing atmosphere \citep{lichtenberg_redox_2021}, while inefficient core differentiation due to an oxidised formation scenario would conversely lead to an oxidising mantle and atmosphere \citep{elkins2008coreless}. 

There are many reasons why interior structure and dynamics are linked; we provide a few more examples here. The existence of a magnetic field --- shielding the surface and atmosphere from radiation --- depends on core size, amongst other factors. A smaller core has been shown to favour a magnetic field on longer timescales due to less efficient cooling of the core through the thick, radiogenically heated mantle \citep{bonati2021structure}. Thicker rocky mantles would permit volcanic outgassing over longer time scales \citep{noack2014can}, at least for low-mass rocky planets. The shallower pressure gradients permit deeper melt zones and facilitate melt extraction from deeper within the mantle. In addition, the cooling of thick mantles is less efficient, so that these planets retain their heat for a longer time. For more massive planets, volcanic activity depends on additional factors including the interior energy budget and the composition and compressibility of the melt, which may not be able to rise to the surface on massive rocky planets \citep{ohtani1995melting}. Consideration of this melt buoyancy effect has lead to different predictions for volcanic activity on super-Earths \citep{kite2009geodynamics,noack2017volcanism}. For example, \citet{noack2017volcanism} find that volcanic activity becomes strongly limited at planet masses above 5~\ME. Similarly, predictions of the likelihood of plate tectonics on super-Earth vary depending on other interior properties such as heating rate or mantle composition (an in-depth literature overview can be found in \citealp{ballmer2021diversity}). More recent studies have started to investigate the effects of variable mantle compositions (e.g. iron fraction as well as Mg/Si content) on mantle geodynamics \citep{dorn2018outgassing,spaargaren2020influence}: silicate composition influences mantle rheology and viscosity, melting temperatures, and other driving mantle properties. All of these studies require up-to-date interior structure models as a first step.

\section{Conclusion and perspectives}

As seen in this review, determining the interior structure of rocky exoplanets is a complex multifaceted problem. Our interpretations of observations rely on indirect data alongside assumptions drawn from our knowledge of the Solar System. Many of the parameters that influence the interior remain unknown, difficult to measure, or potentially entirely unknowable. Compared to the wealth of information we have for the Solar System, the data available for exoplanets is limited and will likely remain so for the foreseeable future.

However, what exoplanet research may lack in terms of data diversity, it makes up for in scope: The growing population of detected, potentially rocky, exoplanets now far exceeds the small number of terrestrial planets in the Solar System. This allow for the first time the use of a statistical approach to planetary science testing the hypotheses about planet formation and evolution that the one data point of our Solar System can not provide. Furthermore, the discovery of planets with no analogue in the solar system present an opportunity to expand our knowledge and test different theories of formation. 

We have entered the age of exoplanet atmosphere characterization owing to space telescopes such as the JWST and the upcoming ARIEL mission. These data can significantly contribute to deepen our understanding of exoplanets and their bulk and interior composition. Atmospheres on rocky planets are shaped by their interaction with the geodynamic processes taking place in the interiors. As atmosphere and interior are part of a single complex co-evolving system, the planets we observe today are the product of billions of years of evolution. To this end, one of the main scientific goals of the soon to be launched PLATO mission is to determine precise host star ages, and therefore the ages of their planets.

Understanding the interior-atmosphere link, and its time evolution, will be one of the major challenges in the upcoming years, but can lead to great rewards when coupled with the developments in observational techniques and instrumentation, improvements to numerical models, and upcoming space missions.
These advancements will need to be combined with enhanced thermodynamic and modelling frameworks able to capture the evolution of the coupled interior-atmosphere system. Necessary to the development of such frameworks is the collection of high-quality datasets constraining the physicochemical properties of materials at high pressure and temperature. More efforts should be devoted to calibrate thermodynamic databases at extreme pressure and temperatures, and implement them with new mineralogical phases that might be uncommon on Earth but can exist for other formation conditions.  

As our observational, experimental, and theoretical tools continue to improve, we move away from treating planets as just static spheres defined by bulk properties towards planets as dynamic evolving systems. Ultimately, determining the interior structure of rocky exoplanets provides the foundation on which all subsequent geophysical and atmospheric research builds upon. The main challenges ahead will involve linking diverse planetary processes, such as mantle convection, tectonics, volcanic outgassing, and the evolution of the atmosphere and surface. Perhaps most importantly, we should foster more collaboration among the expert communities studying each of these areas. Interpreting exoplanet observations will require a holistic understanding of planetary systems that no single community possesses in isolation. Therefore, true progress in our understanding of rocky exoplanets will require a strong, multidisciplinary approach.

\backmatter

\bmhead{Acknowledgements}
The authors thank the two anonymous reviewers whose comments helped to improve the quality of this manuscript, as well as to the International Space Science Institute and the organizers for hosting and managing the workshop "The Geoscience of (Exo)planets: Going beyond habitability", from which this article originated. P.B., A.T., and L.N. are funded by the European Union (ERC, DIVERSE, 101087755). Views and opinions expressed are however those of the author(s) only and do not necessarily reflect those of the European Union or the European Research Council Executive Agency. Neither the European Union nor the granting authority can be held responsible for them. F.M. acknowledges support from the Carnegie Institution for Science and the Alfred P. Sloan Foundation under grant G202114194. CMG is supported by the UK STFC [grant number ST/W000903/1]. EB and AR acknowledge the financial support of the SNSF (grant number: 200021\_197176 and 200020\_215760).
This work has been carried out within the framework of the NCCR PlanetS supported by the Swiss National Science Foundation under grants 51NF40\_182901 and 51NF40\_205606.

\section*{Declarations}

\begin{itemize}
\item The authors declare no competing interests.
\end{itemize}


\bibliography{sn-bibliography}

\end{document}